SPECTRAL PROPERTIES OF LARGE GRADUAL SOLAR ENERGETIC PARTICLE

EVENTS – II – SYSTEMATIC Q/M-DEPENDENCE OF HEAVY ION SPECTRAL BREAKS

M. I. DESAI[1,2], G. M. MASON[3], M. A. DAYEH[1], R. W. EBERT[1], D. J. McCOMAS,[1,2], G. Li[4], C. M. S. COHEN[5], R. A. MEWALDT[5], N. A. SCHWADRON[6,1] AND C. W. SMITH[6]

[1]Southwest Research Institute, 6220 Culebra Road, San Antonio, TX 78238, mdesai@swri.edu

[2]Department of Physics and Astronomy, University of Texas at San Antonio, San Antonio, TX 78249, USA

[3]Johns Hopkins University / Applied Physics Laboratory, Laurel, MD 20723

[4]CSPAR, University of Alabama in Huntsville, Huntsville, AL 35756

[5]California Institute of Technology, Pasadena, CA 91125

[6]University of New Hampshire, 8 College Road, Durham NH 03824

SHORT TITLE: SEP SPECTRAL BREAKS

# ABSTRACT

We fit the ~0.1-500 MeV nucleon[-1] H-Fe spectra in 46 large SEP events surveyed by Desai et al. (2015) with the double power law Band function to obtain a normalization constant, low- and high-energy parameters $\gamma_a$ and $\gamma_b$; and break energy $E_B$. We also calculate the low-energy power-law spectral slope $\gamma_1$. We find that: 1) $\gamma_a$, $\gamma_1$, and $\gamma_b$ are species-independent, and the spectra steepen with increasing energy; 2) $E_B$'s are well ordered by Q/M ratio, and decrease systematically with decreasing Q/M, scaling as $(Q/M)^\alpha$ with $\alpha$ varying between ~0.2-3; 3) $\alpha$ is well correlated with Fe/O at ~0.16-0.23 MeV nucleon[-1], but not with the ~15-21 MeV nucleon[-1] Fe/O and the ~0.5-2.0 MeV nucleon[-1] $^3He/^4He$ ratios; 4) In most events: $\alpha<1.4$, the spectra steepen significantly at higher energy with $\gamma_b-\gamma_a>3$, and O $E_B$ increases with $\gamma_b-\gamma_a$; and 5) Many extreme events (associated with faster CMEs and GLEs) are Fe-rich and $^3He$-rich, have large $\alpha\geq1.4$, flatter spectra at low and high energies with $\gamma_b-\gamma_a<3$, and $E_B$ that anti-correlates with $\gamma_b-\gamma_a$. The species-independence of $\gamma_a$, $\gamma_1$, and $\gamma_b$ and the systematic Q/M dependence of $E_B$ within an event, as well as the range of values for $\alpha$ suggest that the formation of double power laws in SEP events occurs primarily due to diffusive acceleration at near-Sun CME shocks and not due to scattering in the interplanetary turbulence. In most events, the Q/M-dependence of $E_B$ is consistent with the equal diffusion coefficient condition while the event-to-event variations in $\alpha$ may be driven by differences in the near-shock wave intensity spectra, which are flatter than the Kolmogorov turbulence spectrum but weaker than for extreme events. The weaker turbulence allows SEPs to escape easily, resulting in weaker Q/M-dependence of $E_B$, (lower $\alpha$ values) and spectral steepening at higher energies. In extreme events, the flatter spectra at high and low energy and stronger Q/M-dependence of $E_B$ (larger $\alpha$ values) occur due to enhanced wave power, which also enables the faster CME shocks to accelerate flare suprathermals more efficiently than ambient coronal ions.

Subject headings: acceleration of particles --- interplanetary medium --- shock waves --- solar wind --- sun: abundances --- sun: flares

# 1. INTRODUCTION

Large gradual solar energetic particle (SEP) events are believed to be accelerated via diffusive shock acceleration (DSA) mechanisms at shock waves driven by fast coronal mass ejections (CMEs) that plough through the solar corona and interplanetary medium (e.g., Reames, 1999; 2013; Lee 2005; Desai & Giacalone 2015). Such large SEP events, if sufficiently intense, can significantly increase radiation levels in the near-Earth environment, thus damaging technological assets and adversely affecting the health and safety of humans in space (e.g., Desai & Giacalone 2015). Previous studies have shown that the differential energy spectra of H-Fe nuclei in large SEP events exhibit two distinct (or broken) power-laws above and below a characteristic roll-over or break energy, with the break energy typically decreasing for the heavier ion species, or more precisely, with the ion's charge-to-mass or Q/M ratio (e.g., McGuire von Rosenvinge & McDonald 1984; Ellison & Ramaty, 1985; Mazur et al., 1992; Mewaldt et al., 2012). Mewaldt et al. (2005a) suggested that this systematic Q/M-dependence occurs because the energy spectra, usually plotted in MeV nucleon$^{-1}$, steepen or roll over at the same value of the diffusion coefficient for different species, which depends on ion rigidity or the M/Q ratio (see Tylka et al., 2000; Cohen et al., 2005; Mewaldt et al., 2005a).

Indeed, in surveying the Fe and O spectral properties during 46 isolated, large gradual SEP events observed in solar cycles 23 and 24, Desai et al. (2015; hereafter also referred to as Paper

1) found that the Fe spectra had lower break energies owing to the lower Q/M ratio or higher rigidity of Fe when compared with O. Furthermore, Mewaldt et al. (2005a) reported that the observed Q/M-dependence of the spectral break energies during the October 2003 SEP/interplanetary shock event scaled as $(Q/M)^\alpha$ with α≈1.75. This value is smaller than the α≈2 predicted by Li, Zank & Rice (2005) for quasi-parallel shocks, but larger than the ~1.5-1.6 predicted by Battarbee, Laitien & Vainio (2011). Later, Li et al. (2009) generalized their SEP acceleration model by including different levels and slopes for the turbulence spectra at shocks with different obliquity and predicted that α could range between ~0.2 for weaker scattering near quasi-perpendicular shocks and ~2 for stronger Q/M-dependent scattering near quasi-parallel shocks. More recently, Schwadron et al. (2015a;b) developed a new SEP acceleration model where double power-laws occur naturally from shocks and compressions low in the corona, particularly on the flanks of CME expansion regions. In this model, the finite size of the CME shock and stronger Q/M-dependence of the diffusion coefficient facilitates particle escape from the acceleration region, which reduces the break energy and steepens the higher energy spectrum. Conversely, in the Schwadron et al. model, a weaker Q/M-dependence inhibits particle escape, which increases the break energy and flattens the higher energy spectrum. In this paper, we fit the ~0.1-500 MeV nucleon[-1] H-Fe spectra in the 46 large SEP events surveyed in Paper 1 with the Band function to obtain a normalization constant, low- and high-energy parameters $\gamma_a$ and $\gamma_b$; and break energy $E_B$. For each SEP event, we fit the break energy $E_X$ of each species X, normalized to that of H $E_B$ ($E_H$) with $E_X/E_H \propto (Q_X/M_X)^\alpha$, and then investigate properties of α. We compare our results with existing and evolving SEP acceleration models to better understand the physical mechanisms that may be responsible for producing the double power-law spectral forms in large SEP events and the Q/M-dependence of the break energies.

## 2. INSTRUMENTATION & DATA ANALYSES

We use energetic ion data from (1) the Ultra-Low-Energy Isotope Spectrometer (ULEIS: Mason et al., 1998), (2) the Solar Isotope Spetcrometer (SIS: Stone et al., 1998a), and the Electron, proton, and alpha monitor (EPAM: Gold et al., 1998) on board NASA's Advanced Composition Explorer (ACE: Stone et al., 1998b) launched in 1997 August. We also use proton data from the Proton and Electron Telescope on board the Solar, Anomalous, and Magnetospheric Particle Explorer (PET: Cook et al., 1993), the Energetic and Relativistic Nuclei and Electron experiment (ERNE: Torsti et al., 1995) on board the joint ESA/NASA Solar and Heliospheric Observatory (SoHO), and the Energetic Particle Sensor (EPS) on NOAA's Geostationary Operational Environmental Satellites (GOES, series 8-15). Details of these instruments and their species and energy coverage are provided in Table 1.

In Paper 1, we described our event selection criteria and method for identifying sampling intervals for 46 isolated, large SEP events observed at 1 AU from November 1997 through April 2014. None of these SEP events were accompanied by local interplanetary-shock accelerated energetic storm particle (ESP) populations above ~0.1 MeV nucleon$^{-1}$. Tables 1 & 2 of Paper 1 provide the solar source properties, fluence sampling intervals, the ~0.5-2.0 MeV nucleon$^{-1}$ $^{3}$He/$^{4}$He ratio, and the Fe/O ratios at ~0.16-0.23 MeV nucleon$^{-1}$ and ~15-21 MeV nucleon$^{-1}$ associated with these 46 events. In this study for each SEP event, we used ACE/ULEIS, ACE/SIS, GOES/EPS, SoHO/ERNE, and when available, SAMPEX/PET, to obtain the event-integrated ~0.1–500 MeV nucleon$^{-1}$ fluence spectra for 11 species in the range H-Fe, as shown in the three examples in Figure 1. As in Paper 1, we used the non-linear least-squares Levenberg-Marquardt technique and minimized the $\chi^2$ to fit the four-parameter Band function (see Band et al., 1993; Eq. 1) given by:

$$\frac{dj}{dE} = CE^{-\gamma_a} \exp\left(-\frac{E}{E_B}\right) \text{ for } E_T \leq (\gamma_b - \gamma_a)E_B$$

$$\frac{dj}{dE} = CE^{-\gamma_b} \left[(\gamma_b - \gamma_a)E_B\right]^{\gamma_b - \gamma_a} \exp(\gamma_a - \gamma_b) \text{ for } E_T \geq (\gamma_b - \gamma_a)E_B \quad \text{(Eq. 1)}.$$

Here $C$ is the normalization constant, $\gamma_a$ and $\gamma_b$ are the low-energy and high-energy Band-parameters, and $E$, $E_B$, and $E_T$ are respectively, the kinetic, spectral break, and spectral transition energy measured in MeV nucleon$^{-1}$. For each Band-fit parameter, we obtained the formal 1σ uncertainty from the off-diagonal terms of the covariance matrix (Markwardt, 2009). Like the Fe and O spectra discussed in Paper 1, we found that for most species in most SEP events (see Figures 1a, 1c, and 1e), the fits are visually and statistically reasonable, with reduced $\chi^2$ values having ~50% probabilities (also see Mewaldt et al., 2012).

In Paper 1 we showed that the non-orthognality of the Band function results in strong coupling between the O Band-parameters $\gamma_a$ and $E_B$, and further that $\gamma_a$ can be significantly different from what is commonly described as the low-energy power-law spectral slope $\gamma_1$. In order to obtain a physically meaningful quantity that represents the low-energy portion of the SEP spectra below the break energies more accurately, we calculate the low-energy spectral slope $\gamma_1$ between ~0.1-1 MeV nucleon$^{-1}$ for each species in all SEP events using Eq. 1a and the corresponding Band-parameters $\gamma_a$ and $E_B$. For each event and species, we also investigate the properties of the transition energy $E_T$ given by $(\gamma_b - \gamma_a)*E_B$ from Eq. 1. Figure 2 illustrates the relationships between the various spectral parameters studied in this paper, and how the Band parameters $\gamma_a$ and $\gamma_b$ and the spectral slope $\gamma_1$ change when the corresponding portion of the spectrum flattens or steepens. For instance, $\gamma_a$, $\gamma_b$, and $\gamma_1$ increase when the spectrum steepens, and vice versa.

Mewaldt et al. (2012) fitted the four-parameter Band function to ~0.05-500 MeV proton fluence spectra for 16 SEP events that were associated with Ground Level Enhancements or GLEs during solar cycle 23. Seven of these GLE-associated SEP events are included in our survey (see Tables 1 & 2 in Paper 1, and Table 2 here). For these events, we compared the proton fluences obtained from our analyses with those published in the Mewaldt et al. study. In general, the proton fluence spectra from the two independent surveys for all 7 SEP events are in excellent agreement within the stated ~20% uncertainties that account for differences between various instruments: GOES/EPS, SoHO/ERNE, SAMPEX/PET, ACE/EPAM, and ACE/ULEIS. Figure 1e shows the H-Fe fluence spectra during the 2001 December 26 GLE-associated SEP event; here the proton fluences, the corresponding fits, and the fit parameters from the two surveys are nearly identical.

Table 2 provides detailed information about the 46 SEP events: Column (1): Event Number; Column (2) Year; Column (3): SEP fluence sampling interval; Columns (4)-(6): H Band-parameters; Columns (7)-(9): O Band-parameters; Columns (10)-(12): Fe Band-parameters. Column (13): lists the power law slopes of the species-dependent spectral break energies that are obtained as follows. As shown for three SEP events in Figure 1, we obtain and fit the event-integrated fluences for 11 species H, He, C, N, O, Ne, Mg, Si, S, Ca and Fe with the Band function (Eq. 1). For each SEP event, we fitted the roll-over or break energy $E_X$ of each species X normalized to the proton Band-spectral break energy, $E_H$ with a power-law of the form $log(E_X/E_H)=n_0*log[(Q_X/M_X)]^{\alpha}$. Column (13) of Table 2 lists the value of $\alpha$ for each event (also see §3.5). Examples of three different types of Q/M-dependence of $E_X/E_H$, i.e., three different values for the power-law exponent $\alpha$, are shown in Figures 1b, 1d, 1f. For the charge states Q we used the average SEP ionic charge state <Q> determined by Möbius et al. (2000) and Klecker, Möbius & Popecki (2007), namely $He^{2+}$, $C^{5.6+}$, $N^{6.6+}$, $O^{6.8+}$, $Ne^{8.2+}$, $Mg^{8.9+}$, $Si^{9.5+}$, $S^{10.2+}$, $Ca^{10.8+}$, and $Fe^{11.6+}$.

## 3. GENERAL PROPERTIES OF THE SPECTRAL FITS

### 3.1 PROPERTIES OF SEP-BAND PARAMETERS $\gamma_a$ AND $\gamma_b$

In Figure 3a we show statistical properties of the SEP spectra by plotting histograms of the SEP Band-parameters $\gamma_a$ (red) and $\gamma_b$ (blue) for all species in all 46 SEP events. Only those values with relative uncertainties <100% and finite values for $\gamma_a$ and $\gamma_b$ are included; 398 spectra were fitted. Consistent with the statistical properties of the O Band-parameters in Paper 1, we note that $\gamma_a$ for all species has a mean $\mu\sim1.23$ and median $m\sim1.19$, which are both significantly smaller compared to the corresponding values for the high-energy Band parameter, $\gamma_b$: $\mu\sim3.63$ and $m\sim3.57$. We also note that $\gamma_a$ has a narrower distribution with a $1\sigma$ standard deviation of $\sim0.58$, and varies over a substantially smaller range of values between $\sim0$-3.5. In contrast, $\gamma_b$ exhibits a broader distribution with $1\sigma$ value of $\sim1.12$, and varies between $\sim0.7$-9.

Figure 3b shows a scatter plot of $\gamma_b$ vs. $\gamma_a$ obtained for each individual species in all SEP events. As seen for the O spectral slopes in Paper 1, we note that most SEP spectra are flatter at lower energies and steepen above the break energy. The energy spectra in 10 cases flatten at higher energies (see Table 2); these are for H: events #30 & #40; He: event #44; O: event #31; Si: events #2 & #34; Ca: events #2 & #32; and Fe: event #34 & #45. Event #31 was discussed in Paper 1. Since $\gamma_b<\gamma_a$ occurs for different species in different events, including these outliers in our analyses does not affect the overall results and conclusions of this paper.

### 3.2 DIFFERENCES IN H, O, AND FE SEP-BAND PAREMETERS

Figure 4 investigates the relationships and differences between the SEP Band-parameters for different species: (a) $\gamma_a$ for H and Fe vs. $\gamma_a$ of O; (b) $\gamma_b$ for H and Fe vs. $\gamma_b$ of O; (c) $\gamma_b$–$\gamma_a$ for H and Fe vs. $\gamma_b$–$\gamma_a$ of O; and (d) $E_B$ for H and Fe vs. $E_B$ of O. In general, the SEP Band-parameters for H and Fe track those of O reasonably well over the corresponding range of values. We note the following: (1) In many events, $\gamma_a$ for H and Fe are different compared to that of O; the differences between H and O are somewhat larger. (2) $\gamma_b$ and $\gamma_b$–$\gamma_a$ for H and Fe show tighter correlations with the corresponding values for O. (3) $E_B$ for H and Fe in most SEP events show significant differences compared to corresponding values for O; the H $E_B$ in 33 out of 40 events (~83%) is larger than that of O, while the Fe $E_B$ in 26 out of 40 events (~65%) is smaller than that of O (also see Desai et al., 2015).

## 3.3 EVENT-TO-EVENT VARIATONS

Figure 5 examines the event-to-event variations of the SEP spectral parameters for all species: (a) $\gamma_1$; (b) $\gamma_b$; and (c) $E_B$. Red dots show the values for each species in each event, the solid black curves in (a) and (b) show the mean value for $\gamma_1$ and $\gamma_b$, respectively, and the solid black curve in (c) shows the proton Band-spectral break energy, $E_H$, for each event. Dotted lines show the species-averaged mean values obtained by averaging for all species in all 46 events; yellow shaded regions depict the 1σ standard deviation (also see Figure 3b). $\gamma_1$ for events #20 and #34; $\gamma_b$ and $E_B$ for event #20 are not plotted (see §3.5). We note the following: (1) $\gamma_1$ has an event-averaged mean of ~1.64±0.03, with 1σ standard deviation of ~0.6; large deviations of >> 1σ from the event-averaged mean value are seen in events #12, #28, #32, #43 and #45, where $\gamma_1 \geq 2.5$ for most species. (2) In most events, $\gamma_b < 6$ for most species; exceptions are events #17 and #18, where $\gamma_b > 6$ for most species; and (3) In most events, the H $E_B$ is greater than those for the heavier species in the same event. Note that in events #30 and #40, H $E_B$ is greater than ~100 MeV. Overall, within an event, the proton spectral parameters $\gamma_1$ and $\gamma_b$ do not stand out from the corresponding heavy ion spectral parameters but rather they lie in the midst of the others; in contrast, the proton spectral break energy is almost always greater than the heavy ion spectral break energy.

## 3.4 SPECIES-DEPENDENT VARIATIONS WITHIN AN EVENT

We now investigate the species-associated variations in the (a) low-energy spectral slope $\gamma_1$ and Band-parameters $\gamma_a$, and (b) $\gamma_b$ within an event by plotting the distributions of their corresponding mean deviations in Figure 6. The mean deviation for each parameter in each event is calculated from the data shown in Figure 5. Figure 6c shows the actual distribution of the Band-spectral break energy $E_B$ for all species in all SEP events compared to that of the transition energy $E_T$. The following features are striking: (1) The mean deviations of $\gamma_1$, $\gamma_a$, and $\gamma_b$ have narrow distributions which results in well-behaved Gaussian-like distributions (black curves) with small $1\sigma$ values. (2) Typically, $E_T > E_B$, with the mean of $E_T$ almost a factor of 2 greater than that of $E_B$. More importantly, $E_B$ and $E_T$ vary over more than three orders of magnitude, which results in broad distributions with large $1\sigma$ standard deviations. These results indicate that, within a given SEP event, the three parameters $\gamma_1$, $\gamma_a$, and $\gamma_b$, have remarkably similar values for all species, that is, for each event both the low-energy and high-energy spectral slopes are the same to within ~10-15%.    Secondly, for a given event, most of the species-associated, spectral variations are driven by differences in the break energy or the transition energy $E_T$ at which the spectra steepen. Three examples of such species-associated differences in $E_B$ were shown in Figures 1b, 1d, and 1f.

## 3.5 EVENT-TO-EVENT VARIATIONS IN THE Q/M-DEPENDENCE OF $E_B$

To investigate the species-dependence of $E_B$ within an SEP event, as seen in Figures 1b, 1d, 1f, 4d, 6c, as well as the event-to-event variations in $E_B$ shown in Figure 5c, we examine the $E_X/E_H$ vs. the ion's Q/M ratio for 38 SEP events in Figure 7.  The figure shows that in most cases the break energies are well ordered by Q/M ratio.  (In the 5 events not shown, either the ULEIS and SIS spectra for many species did not match near overlapping energies (event #20; also see §3.3), or only the proton spectra showed a spectral break (events #29, #30, #41), or the spectra for most species are well-described by a single power-law across the combined ULEIS and SIS energy range (event #38).)

As in Figure 1, we obtain the power-law exponent $\alpha$ for each event; values of $\alpha$ are provided in Table 2 and in each panel. In 4 events, events #4, #9, #19, and #31, the uncertainties in $\alpha>100\%$. In 4 other events, events #2, #3, #7, and #45, $\alpha<0$. From Table 2, we note that in event #7, the Fe $E_B\sim81$ MeV nucleon$^{-1}$ compared with $\sim11$ MeV nucleon$^{-1}$ for both H and O (and for many other species), which results in the Fe $E_B$ being an outlier in Figure 7, resulting in $\alpha<0$. For events #2, #3, and #45 we only used the $>0.2$ MeV nucleon$^{-1}$ He-Fe spectra ($>0.5$ MeV nucleon$^{-1}$ for event #45), because below these energies, the ULEIS spectra for most species exhibited an upturn or downturn such that the Band function could not be fitted to the entire spectrum. When we force-fitted the spectra by including these lower energy data, the resulting Band fits were poor and the Q/M vs. $E_X/E_H$ plots were similar to that shown for event #31, with uncertainties in $\alpha>100\%$. We therefore exclude these 8 outlier events along with the 5 events not shown in Figure 7 from Table 2 and the subsequent discussion concerning $\alpha$.

The figure clearly shows that the fit (solid red line) and the slope $\alpha$ well characterizes the systematic Q/M-dependence of the heavy ion spectral break energies in most of the remaining 33 SEP events in our survey. Other noteworthy features are: (1) In these 33 SEP events, including the 3 cases shown in Figure 1, $\alpha>0.2$ (dashed black line). (2) In events #16, #25, and #32, $\alpha>2$ (dotted black line, also see §4.1). Note that in event #32, the values of $E_X/E_H$ for Ca and Fe are significantly larger than the fit, which yields $\alpha\sim2.16$. Likewise, larger values of $E_X/E_H$ for Fe and/or Ca are also seen for events #21, #28, #37, and #42, resulting in large deviations from the fits. Such deviations are probably affected by our assumption of the average SEP charge states for these species. For instance, if both the Ca and Fe were highly ionized in these events, as has been observed in some large SEP events with significant contributions of flare material in the seed population (e.g., Klecker et al. 2007), then their corresponding Q/M values would be larger. This would shift the data points to the right and closer to the fitted line. Indeed, Table 3 shows that four out of these five events have enrichments in the $^3$He/$^4$He ratio and in the Fe/O ratio when compared to the corresponding abundances measured in the solar wind, indicating the presence of flare-rich suprathermals in the seed population.

## 4. GENERAL PROPERTIES OF THE POWER-LAW EXPONENT $\alpha$

### 4.1 DISTRIBUTIONS AND RELATIONSHIP WITH SOLAR SOURCE PROPERTIES

We now investigate the properties of $\alpha$ and its relationship with solar source properties given in Table 1 of Paper 1. Figure 8a shows the histogram of $\alpha$ along with the mean, standard deviation, and median value of the distribution, while Figures 8b, 8c, and 8d plot $\alpha$ vs. flare longitude, CME speed, and the peak proton flux (PFU) obtained by NOAA GOES/EPS, respectively. Figure 8 shows the following: (1) $\alpha$ has a mean value of 1.27, median value of 1.16, and is confined between ~0.2–3, with values for three SEP events greater than 2; two of these events were also accompanied by GLEs (see Table 2, Figures 1f and 6); (2) $\alpha$ exhibits no clear trend with the flare longitude, but SEP events with $\alpha>2$ are associated with source longitudes west of W45; note that this is probably a result of the selection criteria which is biased toward western heremisphere events; and (3) $\alpha$ exhibits statistically significant, positive trends with the peak proton flux and CME speed, with values for correlation coefficients of r~0.48 and r~0.41, which have probabilities of <1% and <2%, respectively of being exceeded by uncorrelated pairs of parameters. It is evident from the figure that the correlations with CME speed and peak proton flux are largely due to the presence of events with high CME speeds and/or GLE events.

## 4.2 RELATIONSHIP BETWEEN α AND SEP FE/O AND $^3$HE/$^4$HE RATIOS

Figure 9 examines the relationships between α and key heavy ion abundances: (a) the ~0.16–0.23 MeV nucleon$^{-1}$ Fe/O, (b) the ~15–21 MeV nucleon$^{-1}$ Fe/O, and (c) the ~0.5–2.0 MeV nucleon$^{-1}$ $^3$He/$^4$He ratios; all three ratios are taken from Table 2 in Paper 1. The ~0.16–0.23 MeV nucleon$^{-1}$ Fe/O exhibits a statistically significant, positive trend with α; correlation coefficient $r$~0.44 for 31 events has <1% chance of being exceeded by an uncorrelated pair of parameters. In contrast, α is not well correlated either with the ~15–21 MeV nucleon$^{-1}$ Fe/O or the ~0.5–2.0 MeV nucleon$^{-1}$ $^3$He/$^4$He ratios.

## 4.3 RELATIONSHIP BETWEEN α AND O BAND-PARAMETERS $\gamma_a$ and $\gamma_b$

Figure 10a investigates the relationship between the O SEP Band-parameters $\gamma_a$ and $\gamma_b$ and the power-law exponent α, while Figure 10b plots the difference, $\gamma_b - \gamma_a$ vs. α. We note the following: (1) α is not correlated with $\gamma_a$, $\gamma_b$, and $\gamma_b - \gamma_a$; and (2) SEP events with α between ~0.6-1.4 have a larger range of values for $\gamma_b$ and $\gamma_b - \gamma_a$, i.e., events for which the spectra steepen significantly at higher energies occur for α values between ~0.6-1.4.

Figure 11 examines the relationship between the O break energy $E_B$ and (a) the difference between the spectral slopes $\gamma_b - \gamma_a$, and (b) α. Overall, the O $E_B$ is not correlated with $\gamma_b - \gamma_a$ or α. However, the O $E_B$ in SEP events with $\gamma_b - \gamma_a < 3$ and $\gamma_b - \gamma_a > 3$ appear to exhibit negative and positive trends, respectively.

# 5. DISCUSSION

Paper 1 investigated properties of the Fe and O fluence spectra in 46 isolated, large gradual SEP events observed at 1 AU during solar cycles 23 and 24. In this paper, we fit the event-integrated fluence spectra of ~0.1-500 MeV nucleon[-1] H-Fe in the same 46 SEP events with the four-parameter Band function and investigate properties of the SEP Band-parameters $\gamma_a$, $\gamma_b$, and $E_B$.  We also calculate the low-energy power-law spectral slope $\gamma_1$. Our results are:

1)   Figure 3 in §3.1 shows that $\gamma_a$ ranges between ~0.1-3 and has a mean value $\mu$~1.2; $\gamma_b$ ranges between 0.5-9 and has $\mu$~3.6. $\gamma_a$ is also typically smaller than $\gamma_b$, implying that the energy spectrum of each species in any given SEP event steepens with increasing energy.

2)   Figures 4a-c in §3.2 and 4a-b in §3.2 show that in most SEP events, $\gamma_a$, $\gamma_1$, $\gamma_b$, and $E_B$ for different ion species track each other well. Figures 6a-b in §3.3 show that, within a given SEP event, $\gamma_a$, $\gamma_b$, and $\gamma_1$ for H-Fe are nearly identical, each mean deviation exhibits a Gaussian-like distribution with a small 1$\sigma$ of $\lesssim$0.08.

3)   Figure 4d in §3.2 shows that, in general, $E_H > E_O > E_{Fe}$. Figure 5c in §3.3 shows that, in most SEP events, $E_H$ generally exceeds the $E_B$ for the heavier ion species. Figure 6 shows that species-dependent variations in $E_B$'s and transition energies $E_T$'s occur over three orders of magnitude.

4)   Figures 1b, 1d, 1f and 6 show that $E_B$'s in 33 of the 46 SEP events in our survey vary systematically according to the ion's Q/M ratio, and further that the event-to-event power-law slope of this dependence can be characterized by a single parameter $\alpha$–given by fitting $log(E_X/E_H)=n_0*log[(Q_X/M_X)]^{\alpha}$ for each event.

5)   Figure 8 in §4.1 shows that for 33 SEP events, $\alpha$ varies between ~0.2-3. $\alpha$ exhibits statistically significant, positive trends with the peak proton flux and CME speed.

6)      Figure 9 in §4.2 shows that α is positively correlated with the Fe/O ratio at ~0.16-0.23 MeV nucleon$^{-1}$, but not with the ~15-21 MeV nucleon$^{-1}$ Fe/O and the ~0.5-2.0 MeV nucleon$^{-1}$ $^{3}$He/$^{4}$He ratios.

7)      Figure 10 in §4.3 shows that α is not correlated with $\gamma_a$, $\gamma_b$, and $\gamma_b-\gamma_a$. Events with α<0.6 and α ≥1.4 also have low values for $\gamma_a$, $\gamma_b$, and $\gamma_b-\gamma_a$. Events with 0.6<α<1.4 have a larger range of values for $\gamma_b-\gamma_a$.

8)      Figure 11 in §4.3 shows that the O break energy $E_B$ is not correlated with $\gamma_b-\gamma_a$ or α, but $E_B$ does exhibit a weak negative trend for the group of events in which $\gamma_b-\gamma_a$<3. In contrast, SEP events with $\gamma_b-\gamma_a$>3 exhibit a positive trend between $E_B$ and $\gamma_b-\gamma_a$.

## 5.1 PROPERTIES OF NEAR-SUN CME SHOCKS AND TURBULENCE CONDITIONS

Comparing our survey to prior studies, we note that some of the 5 events studied by Mewaldt et al. (2005a) and Cohen et al. (2005) also included the local shock-accelerated ESP component that accompanied the larger SEP event. In contrast, we had eliminated all events with possible contributions from local IP shock-associated populations (see Paper 1). Further, we use event-integrated fluences, rather than time-intensity profiles (see Mason et al., 2012), to study the SEP spectral properties. In particular, Mason et al. (2012) used a detailed model of interplanetary propagation and showed that transport from the inner solar system can lower the break energy systematically for all species, as well as lower the slopes by 10-20% but that the basic spectral form remained intact (their Fig. 14). Alternatively, we note that Li & Lee (2015) fitted the double power-law proton spectra in 9 of the 16 GLE events studied by Mewaldt et al. (2005a) with an analytical model that included interplanetary transport effects, and found that single power-law spectra injected by CME shocks near the Sun can exhibit spectral breaks at 1 AU due to scatter-dominated transport through the interplanetary medium. However, this model predicts that α for the GLE-associated SEP events range between ~0.18-0.75, which is clearly inconsistent with the α>1.58 observed in 5 of the 7 GLEs in our survey (see Table 2). On this basis, we contend that the formation of the double power law SEP spectral forms, their associated properties, and the observed Q/M-dependence of $E_B$ primarily reflect conditions near the distant CME-driven shocks where the acceleration takes place, and are not significantly affected by contributions from local interplanetary shock-accelerated populations nor by Q/M-dependent transport and scattering in the interplanetary turbulence en route to 1 AU (e.g., see Zank, Rice & Wu 2000; Cohen et al., 2005; Mason et al., 2012).

A fundamental prediction of early 1-dimensional (1D) steady-state as well as the more recent time-dependent DSA-based SEP models is that, in a given event, the differential energy spectrum of the accelerated particles below the break energy is characterized by a low-energy power-law spectral slope γ given by $dj/dE \propto E^{-\gamma}$ (e.g., Drury 1983; Lee 2005, Schwadron et al., 2015b). These models also predict that γ is independent of ion species, and is determined solely by $\gamma \approx (H + 2)/(2H - 2)$, where $H$ is the strength of the CME-driven shock. Our results show that both, the SEP Band-parameter $\gamma_a$ and the low-energy spectral slope $\gamma_1$ in a given SEP event are remarkably similar for all species, and that such species-independent spectral slopes are observed at both low and high energies for most of the events in our survey (Result #'s 1-2). We therefore suggest that, to first order, the formation of double power-law spectra in large SEP events is consistent with DSA at near-Sun CME shocks (e.g., Schwadron et al., 2015b).

We now use the DSA-predicted relationship between $\gamma$ (here we use the species-averaged $\gamma_1$ for each event) and $H$ to infer the compression ratios of the near-sun CME shocks. Figure 12 compares these inferred values to three key properties of CMEs and SEPs: (a) CME speed, (b) peak proton flux, and (c) α from Table 2. The main features of this figure are: (1) the inferred shock compression ratios for the 37 events shown here lie between ~1-5.5, with H>4 for 3 events. These values remarkably consistent with the predicted range of values for CME shock compression ratios (see Schwadron et al., 2015b), and well within the constraints of the Rankine-Huognoit discontinuity conditions for the allowable range of ~1-4 and upper limit (<4) for shock compression ratios in space plasmas (e.g., Viñas & Scudder 1986). We note that all cases of H>4 have sizeable uncertainties.  (2) The compression ratio H exhibits weak but positive correlations with all three parameters: (a) CME speed: for 37 events, r~0.38 has <2% chance; (b) peak proton flux: for 29 events, r~0.35 has <6% chance; and (c) α: for 31 events, excluding events with H>4, r~0.47 has <0.57% chance, of being exceeded by uncorrelated pairs of parameters.

The heavy ion fluence spectra in most SEP events are flat at energies below ~1 MeV nucleon$^{-1}$ and steepen above a roll-over or break energy, which decreases systematically with the ion's Q/M ratio (Results #3 and #4). The Q/M-dependence of $E_B$'s in a given SEP event is well represented by the function $log(E_X/E_H)=n_0*log[(Q_X/M_X)]^\alpha$ and characterized by the power-law exponent α (Result #4). The values of α for 33 SEP events lie in the range ~0.2-3 (Result #5), which encompasses the range of α values found in the surveys of Mewaldt et al. (2005a) and Cohen et al. (2005). Thus, with the exception of 3 events with α>2 (see Figure 7 and §5.2), the range of values for α in our survey is consistent with the corresponding range of ~0.2-2 predicted by Li et al. (2009). In this model, the Q/M-dependence of the spectral break energies in a given SEP event occurs due to the "equal diffusion coefficient" or the "equal acceleration time" condition, and the event-to-event variations in the power-law exponent α are driven by the differences in the slopes of the turbulence spectra that are expected to be present near shocks with different obliquity.

Assuming that the spectral break energies for different species in a given SEP event occur at the same value of the diffusion coefficient, $\kappa_{//}$[1], which scales as $(M/Q)^a$ with a$\approx$0.8–2.7, Cohen et al. (2005) followed Dröge (1994) and inferred that the power-law index $\eta$ of the turbulence or wave intensity spectrum, given by $I(k) \propto k^{-\eta}$, near the CME shock acceleration region, ranged between 1.2 to -0.7. Here $\eta = 2 - a$, and $a$ is related to the exponent $\alpha$ in our survey by $a = \alpha/(2 - \alpha)$. We now follow the approach of Mewaldt et al. (2005a) and Cohen et al. (2005) to infer the power-law exponent $a$, which determines the scaling between the particle diffusion coefficient and the ion's M/Q ratio for 24 SEP events, and the power-law index $\eta$ of the wave intensity spectrum for 27 SEP events in our survey. In this plot, we only include events that satisfied the following criteria: 1) fitted values of $\alpha$ and the inferred values of $a$ and $\eta$ have relative uncertainties <100%, and 2) -4< $\eta$ <+4 (see §5.2).

Figure 13 plots histograms of (a) $a$ and (b) $\eta$; the red histograms represent the extreme SEP events discussed in §5.2. We remark that within the estimated uncertainties, the power-law exponent $a$ in 20 out of 24 SEP events is comparable to those obtained by Cohen et al. (2005), as shown by the yellow shaded region. Also, $a$ varies between ~0.16-3.9, which is roughly consistent with the typical range of ~0.5-7 proposed recently by Schwadron et al. (2015b); in this model a<1 implies weak dependence of $\lambda_{//}$ on the ion's Q/M ratio, while a>1 indicates that $\lambda_{//}$ depends strongly on Q/M (also see Li et al., 2009; Battarbee et al., 2011, 2013; Vainio et al., 2014). We remark that in 7 events, $\eta > 1.2$ – the largest value reported by Cohen et al. (2005); in four of these events $\eta$>5/3. Events with $\eta$>5/3 represent cases in which the turbulence intensity spectra near the distant CME shocks may be significantly steeper than the typical interplanetary Kolmogorov $k^{-5/3}$ turbulence spectrum. In contrast, $\eta$ in 20 out of 27 SEP events lies within the range of values reported by Cohen et al. (2005). In 19 of these events, $\eta$ is less than 5/3, which implies that the turbulence spectra near the corresponding CME shocks are probably significantly flatter than the Kolmogorov spectrum. Finally, result #8 indicates that when $E_B$ is plotted vs. $\gamma_b-\gamma_a$, the SEP events separate into two groups ($\gamma_b-\gamma_a<3$ and $>3$), perhaps indicating that two competing mechanisms are occurring simultaneously in all SEP events: 1) Q/M-dependent processes that produce modest values for $\alpha$ ($<1.4$), and steeper spectra at higher energies that steepen significantly as the break energy increases, and 2) much stronger Q/M-dependent processes that produce higher values of $\alpha$, relatively flatter spectra at high and low energies, and higher break energies (see §5.2).

---

[1] Here $\kappa_{//} = 1/3 v \lambda_{//}$, where $v$ is particle speed and $\lambda_{//}$ is the parallel scattering mean free path.

## 5.2 EXTREME SEP EVENTS

Nine events in our survey can be considered "extreme" events since they produced GLEs (see Mewaldt et al. 2012) or had CME speeds $>2000$ km s$^{-1}$. These events are shown with solid color-coded symbols in Figures 8-12. Taking these events together as a group and comparing with the remaining events in our survey, the extreme events had:

a) stronger dependence of the break energy on Q/M ratio, resulting in $\alpha \geq 1.4$ vs. $\alpha<1.4$;

b) higher peak proton fluxes between $\sim 4\times10^1–2\times10^3$ vs. $\sim10^1–5\times10^2$;

c) source locations from the "well-connected" region of the western hemisphere (longitude locations between W45 – W90 vs. E90 – W120);

d) a stronger positive correlation between the low energy Fe/O and $\alpha$;

e) low-energy spectral parameters $\gamma_a$ and $\gamma_1$ similar to other events, and to the mean and median values of the overall distributions (e.g., $\gamma_a\sim1.2$);

f) flatter spectra at higher energies compared to other events ($\gamma_b \sim2.5–4$ vs. $\sim2.5–7$); and

g)  higher average O break energy compared with other events (~1-12 MeV nucleon$^{-1}$ vs. 0.1- 10 MeV nucleon$^{-1}$).

While some of these features are related to our selection (e.g., faster CME speeds, higher proton flux, flatter high energy spectra), the others are not, and so may provide clues to the properties of extreme SEP events. Figures 10 and 11 show that all of these extreme events with $\alpha \geq 1.4$ have low values for $\gamma_a$, $\gamma_b$, and $\gamma_b - \gamma_a$, and, they collectively exhibit a negative trend between $E_B$ and $\gamma_b - \gamma_a$ (Results #7 and #8). This indicates that the corresponding spectra are relatively flat with similar spectral slopes at low and high energies, and that the break energy increases as the difference $\gamma_b - \gamma_a$ between the SEP O Band-parameters decreases. The fact that $\alpha > 2$ in 2 of the 7 GLE-associated SEP events in our study taken together with the general result that higher values of $\alpha$ ($\geq 1.4$) are typically observed in SEP events that are also associated with higher >10 MeV proton fluxes, faster (>2000 km s$^{-1}$), western hemisphere CMEs, and with GLEs (see Figure 10), indicates that spectral properties in these extreme SEP events are most likely governed by highly efficient trapping and stronger Q/M-dependent scattering due to *substantially enhanced wave power* near the distant CME-driven shocks (see also Cohen et al., 2005; Li et al., 2009).

The above scenario is consistent with the following inferred results shown in this paper:

1)      Figure 12 shows that the inferred values for the shock compression ratio H in these extreme events tend to be somewhat larger than the event average, and that the correlations between $H$ vs. peak proton flux and $\alpha$ are more significant for and therefore likely to be driven by these events.

2)      Figure 13 (red histograms: see §5.1 for the selection criteria we used to infer a and $\eta$) shows that $\lambda_{//}$ has a strong Q/M-dependence with a$\geq$2.5 in 3 of these extreme events, and $\eta > 2$ or $\eta < 1.5$ in 5 of these events, which corresponds to substantially enhanced wave power.

3)      Four of the five SEP events with the strongest observed Q/M-dependence in $E_B$'s, i.e., with $\alpha > 1.6$ and $|\eta| > 4$, are also extreme events, as defined above.

In summary, the extreme SEP events in our survey exhibit strong Q/M-dependencies in the $E_B$'s, and so they have larger values for $\alpha$, which correspond to extreme or above average values for $a$, $\eta$, and $H$.  In contrast, most events in our survey (see §5.1) exhibit weaker Q/M-dependencies in the $E_B$'s and are associated with steeper spectra at higher energies probably because the SEPs are accelerated at much slower and relatively weaker CME shocks, where the somewhat weaker turbulence allows the accelerated particles to escape more easily. Note that this interpretation is at odds with the model of Schwadron et al. (2015a;b), where stronger Q/M-dependence of the diffusion coefficient facilitates particle escape and therefore produces steeper spectra at higher energies.

The question is what special conditions or physical processes are responsible for causing the significantly stronger Q/M-dependence in the spectral break energies in extreme SEP events? We note that in many SEP acceleration models, strong Q/M-dependent scattering occurs primarily at quasi-parallel shocks where turbulence levels are expected to be higher and self-generated Alfvén waves may also be present (Ng, Reames & Tylka 2003; Li et al., 2009). We rule out the possibility that the 5 SEP events with $\alpha > 1.6$ and $|\eta| > 4$ are due to strong scattering at the Bohm diffusion limit, i.e., when $\lambda_{//} \sim \rho_g$, here $\rho_g$ is the ion gyroradius, because the Bohm approximation represents the case for $\eta=1$ and $\alpha=1$ (for details see Li et al., 2009), which is significantly smaller than the $\alpha$ values obtained for these events. Another possible source of enhanced turbulence are the Alfvén waves excited by protons, accelerated at earlier times, that escape and stream upstream of the CME shock (e.g., Zank, Li & Verkhoglyadova 2007). Such self-generated Alfvén waves can trap and scatter the particles that are accelerated at later times much more efficiently (e.g., Ng et al., 2003). Indeed, Ng et al. (2003) modeled the excitation of Alfvén waves by protons streaming away from CME shocks, and found that the wave spectra could exhibit flat spectra with $\eta \approx 0$, which corresponds to a=2 and $\alpha$=4/3. In contrast, Battarbee et al. (2011) modeled SEP acceleration through self-generated turbulence near CME shocks with speeds of 1250, 1500, and 1750 km s$^{-1}$ and predicted that the maximum value for $\alpha$ is ~1.5-1.6. Comparing these predictions with the $\alpha$ in the extreme SEP events suggests that such models, even though they include the non-linear effects of self-generated Alfvén waves, still cannot account for the significantly stronger Q/M-dependence in the heavy ion spectral break energies reported here.

Enhanced turbulence conditions could also occur when the so-called "equal resonance condition" is met, as discussed by Li et al. (2003) and Rice et al. (2003). According to Zank et al. (2007) and Li et al (2009), this condition occurs when $\alpha$=2 and $\eta = \pm \infty$ at parallel shocks in the limit of strong turbulence and scattering, i.e., when the wave power or intensity spectrum $I(k)$ approaches a discontinuity. This is a special case of the more general, the equal acceleration time or equal diffusion coefficient condition discussed in §5.1 (see Li et al., 2009). However, we note that the values of $\alpha$ in 2 GLEs (events #16 and #25) are more than 1σ greater than the maximum value of $\alpha \approx 2$, predicted by Li et al. (2005; 2009). Thus, for these events, $\alpha > 2$ corresponds to a<0 or a>8, and $|\eta| > 4$, i.e., where the wave power $I(k)$ becomes essentially discontinuous. This implies that the scattering and trapping of particles near the distant CME shocks in such events is so strong that the Q/M-dependence of the spectral break energies exceeds the limit of the equal resonance condition. We therefore suggest that the larger than predicted values for $\alpha$ in some of the extreme events in our survey indicates that the underlying mechanisms have not yet been fully incorporated in current theoretical models.

We remark that in most existing theoretical models, the strongest Q/M-dependence occurs at quasi-parallel shocks (e.g., Li et al., 2009). Such shocks, with low injection thresholds, are expected to primarily inject and accelerate the low-energy solar wind or ambient coronal ions (e.g., Tylka & Lee 2006; however, see Giacalone 2005). In contrast, and consistent with the results reported in Paper 1, we find that many of the extreme events that exhibit strong Q/M-dependent spectral break energies are also Fe-rich and $^3$He-rich (see Figure 9 and Paper 1). This points to the importance of contributions of suprathermal flare-origin material to the seed populations for fast CME shocks even in cases when turbulence levels are significantly enhanced. We suggest that in such events, the enhanced turbulence traps, injects, and accelerates the higher-energy suprathermals much more efficiently than the co-existing lower-energy solar wind or coronal suprathermal ions. Simultaneously, the equal diffusion coefficient condition causes the spectral break energies to exhibit stronger Q/M-dependence, occasionally exceeding the equal resonance condition limit, as in the case of 2 SEP events that produced GLEs. We therefore suggest that our results can be reconciled with SEP models provided that they include suprathermal flare-origin material as an important component of the seed population that is available for acceleration at near-Sun CME shocks.

## 6. SUMMARY AND CONCLUSIONS

We fit the ~0.1-500 MeV nucleon$^{-1}$ H-Fe ion fluences in 46 isolated, large gradual SEP events observed during solar cycles 23 and 24 and surveyed in Paper 1 with the four-parameter Band function that yields a normalization constant, low- and high-energy Band parameters $\gamma_a$ and $\gamma_b$, and break energy $E_B$. We also calculate the low-energy power-law spectral slope $\gamma_1$. We find that: 1) In a given SEP event, $\gamma_a$, $\gamma_b$, and $\gamma_1$ are remarkably similar for all species, and the energy spectra steepen with increasing energy; 2) The $E_B$'s in a given SEP event vary systematically according to the ion Q/M ratios, and this dependence is characterized by $\alpha$ – given by fitting $log(E_X/E_H) = n_0 * log[(Q_X/M_X)]^\alpha$; 3) $\alpha$ varies between ~0.2-3, and is well-correlated with the Fe/O ratio at ~0.16-0.23 MeV nucleon$^{-1}$, but not with the ~15-21 MeV nucleon$^{-1}$ Fe/O and the ~0.5-2.0 MeV nucleon$^{-1}$ $^3$He/$^4$He ratios; 4) In most SEP events, $\alpha$<1.4, and the spectra steepen significantly at higher energy with $\gamma_b$–$\gamma_a$>3, and $E_B$ increases with increasing $\gamma_b$–$\gamma_a$; and 5) In many extreme SEP events (those associated with >2000 km s$^{-1}$, western hemisphere CMEs and GLEs), the energy spectra are relatively flatter at low and high energies with the difference $\gamma_b$–$\gamma_a$<3, the break energies increase as $\gamma_b - \gamma_a \to 1$, the events have stronger Q/M-dependence in $E_B$'s with $\alpha \geq 1.4$, and are Fe-rich and $^3$He-rich.

Our results have the following implications for current models of SEP acceleration at near-Sun CME shocks. The species-independence of SEP Band parameters and the low-energy spectral slope, and the systematic Q/M dependence of the break energies within an event, as well as the range of values for $\alpha$ suggest that the formation of double power laws in SEP events occurs primarily due to diffusive shock acceleration at near-Sun CME shocks as predicted by Li et al. (2009) and Schwadron et al. (2015b), and not due to scattering in the interplanetary turbulence as predicted by Li & Lee (2014). Remarkably, our results for the low-energy spectral slopes also correspond to a range of values for CME shocks with compression ratios between ~2-4, as predicted by Schwadron et al. (2015b). Furthermore, the systematic Q/M-dependence of the spectral break energies in a given SEP event is consistent with the equal diffusion coefficient condition in which the energy spectra of different heavy ion species roll over at the same value of the diffusion coefficient, as predicted by Li et al. (2009). The event-to-event variations in $\alpha$ occur due to the differences in the power-law slopes of the wave intensity spectra near the distant CME shocks, and may also provide clues about the remote shock's obliquity. In 27 events, the SEPs are accelerated by CME-driven shocks where the relatively weaker turbulence results in weaker Q/M-dependence of the break energies and lower values for α (<1.4). Even though in the majority of these SEP events (19 of 27), the turbulence spectra are flatter than the typical interplanetary Kolmogorov $k^{-5/3}$ turbulence spectrum, the accelerated SEPs can still easily escape from the CME shock, causing the spectra to steepen significantly at higher energies. In contrast, the significantly stronger Q/M-dependence of the break energies, larger values of α, and the relatively flatter spectra at high and low energies occur in 9 extreme SEP events due to extreme values of the turbulence spectral slopes near faster (>2000 km s$^{-1}$) and stronger CME-driven shocks. We suggest that most DSA-based SEP models (e.g., Ng et al., 2003; Batterbee et al., 2011;2013; Schwadron et al., 2015b, and Li et al., 2009) are unable to fully account for spectral properties in extreme SEP events because the substantially enhanced wave power and associated turbulence scatters, traps, injects, and accelerates suprathermal flare-origin material more efficiently than the co-existing ambient coronal or suprathermal solar wind ions.

We are grateful to the members of the Space Physics Group at the University of Maryland and the Johns Hopkins Applied Physics Laboratory (JHU/APL) for the construction of the ULEIS instrument and to members of Space Radiation Laboratory at the California Institute of Technology for the construction of the SIS instrument. We acknowledge use of the NOAA GOES and SoHO/ERNE proton data. Work at SwRI is partially supported by NASA grants NNX13AE07G and NNX13AI75G, NASA contracts NNX10AT75G and NNN06AA01C, and NSF Grants AGS-1135432 and AGS-1460118.  Work at APL was supported by NASA grants NNX13AR20G/115828 and 44A- 1091698.

FIGURE CAPTIONS:

FIGURE 1:  (a, c, e) Event-integrated differential fluences versus energy of ~0.1-500 MeV nucleon$^{-1}$ H-Fe nuclei during three large SEP events. The energy spectra for different species are offset for clarity. Solid lines: fits to the spectra using the Band function (see Eq. 1; Band et al., 1993). Figure 1e: red data points superposed on the blue symbols are proton data from EPAM/ ACE, EPS/GOES-8, and PET/SAMPEX; the dotted red-curve shows the corresponding band-function fit from Mewaldt et al. (2012) study. (b, d, f) Spectral break energy $E_X$ of species X normalized to $E_H$ -- break energy of H vs. the ion's charge-to-mass (Q/M) ratio. Solid line: fit to the data $\frac{E_X}{E_H} = n_0 (Q/M)_X{}^{\propto}$; dashed line: same equation with $\propto = 2$; dotted line: same equation with $\propto = 0.2$; $\propto$ -- power-law dependence of $E_X/E_H$ on $Q_X/M_X$. The ionic charge states, $Q_X$ for each species are taken as the mean Q-state observed in several large SEP events (Mobius et al., 2000; Klecker et al., 2007).

FIGURE 2:  Example of an SEP spectrum defining the various spectral parameters surveyed in this paper (for details see Eq. 1 and §2). The figure also illustrates the relationships between these parameters and how they change when the spectrum flattens or steepens. Spectral parameters $\gamma_a$, $\gamma_b$, and $\gamma_1$ increase when the spectrum steepens, and decrease when the spectrum flattens.

FIGURE 3: (a) Histograms of SEP Band-parameters $\gamma_a$ (red) and $\gamma_b$ (blue) for all species in all 46 events. N=number of spectra fitted; μ=mean ± standard error of the mean; σ=1-sigma standard deviation; m=median values of the distributions. (b) Scatter-plots of $\gamma_a$ vs. $\gamma_b$. Dotted line shows

spectra for which $\gamma_a = \gamma_b$. All parameters with relative uncertainties $\geq 100\%$ are excluded (see Table 2).

FIGURE 4: Scatterplots of SEP Band-parameters of H and Fe vs. O: (a) $\gamma_a$; (b) $\gamma_b$; (c) $\gamma_b - \gamma_a$; and (d) $E_B$. The red circles show H vs. O, and the blue triangles show Fe vs. O. N=number of data points. Solid lines show 1:1 relationships.

FIGURE 5: *Red dots:* value for all species for the (a) low-energy spectral slope $\gamma_1$ (see §3.3 for details); (b) Band-parameter $\gamma_b$; and (c) Band-parameter, $E_B$ plotted vs. the event number. Solid black curves in (a) and (b) shows the mean value for $\gamma_1$ and $\gamma_b$ in each event, respectively, and the solid black curve in (c) shows the proton Band-parameter, $E_H$ in each SEP event.

FIGURE 6: Histograms of mean deviations for (a) Band-parameter $\gamma_a$ (red), and low-energy spectral slope $\gamma_1$ (blue); and (b) Band-parameter, $\gamma_b$ from the average value for each event. (c) Histograms of Band-parameter $E_B$ (red) and transition energy $E_T$ (blue). N=number of data points; m=median; $\mu$=mean and standard error of the mean; $\sigma$=1 standard deviation of the distribution. The solid black curves show Gaussian fits, with mean and 1$\sigma$ standard deviation, for the distributions of $\gamma_1$ in (a) and $\gamma_b$.

FIGURE 7: Same as Figures 1b, 1d, 1f, but for the remaining 38 of the 41 SEP events that exhibited finite heavy ion spectral breaks that allowed determination of the exponent $\alpha$. The fitted values of $\alpha$ are in Table 2.

FIGURE 8: (a) Histogram of α, and α vs. (b) Flare longitude, (c) CME speed (km s$^{-1}$), and (d) Peak Proton Flux. The peak proton flux, flare longitude, and CME speeds are taken from Table 1 of Paper 1 (Desai et al., 2015). N=number of events plotted; μ, σ, and m as defined in Figure 2; r=correlation coefficient; p=probability that the absolute value of r can be exceeded by an uncorrelated pair of parameters. Green: GLEs; red: CMEs with speeds >2000 km s$^{-1}$; blue: GLEs and CMEs with speeds >2000 km s$^{-1}$.

FIGURE 9: Scatterplots of α vs. (a) 0.16-0.23 MeV nucleon$^{-1}$ Fe/O; (b) 15-21 MeV nucleon$^{-1}$ Fe/O; and (c) 0.5-2.0 MeV nucleon$^{-1}$ $^3$He/$^4$He ratio. All abundance ratios are taken from Table 2 of Paper 1 (Desai et al., 2015). Dashed lines show Fe/O ratios at =0.404 and =0.134, which are average values in several large SEP events at 0.32-0.5 MeV nucleon$^{-1}$ (Desai et al., 2006) and at ~5-12 MeV nucleon$^{-1}$ (Reames, 2013), respectively. Color-coded symbols denote SEP events associated with fast CME and GLES, as in Figure 7.

Figure 10: Scatterplots of α vs. (a) O Band-parameters $\gamma_a$ (red) and $\gamma_b$ (blue), and (b) the difference $\gamma_b$-$\gamma_a$. Color-coded symbols denote SEP events as in Figure 7.

Figure 11: Scatterplots of O Band-parameter $E_B$ vs. (a) the difference $\gamma_b$-$\gamma_a$, and (b) α. Color-coded symbols denote SEP events as shown in Figure 7.

FIGURE 12: Scatterplots of the inferred shock compression ratio vs. (a) CME speed, (b) Peak proton flux, and (c) α. Color-coded symbols denote SEP events as shown in Figure 7. Dotted line shows the H=4 limit for the compression ratio for space plasma shocks. Orange shaded region

encompasses the mean value (solid line), not including the extreme events (see §5.2) and the standard error of the mean.

Figure 13: Histograms of (a) a, the power-law exponent of the dependence of the scattering mean free path $\lambda_{//} \propto (M/Q)^a$, (b) $\eta$, the power-law exponent of the wave intensity spectrum $I$ near the CME-driven shock given by $I \propto k^{-\eta}$. Shaded yellow region: range of values obtained by Cohen et al., (2005) in 5 SEP events; blue vertical line: represents $\eta=5/3$ – the typical interplanetary Kolmogorov spectrum; red histograms represent the extreme SEP events discussed in §5.2.

| Table 1: Data Sources used in this work | | | |
|---|---|---|---|
| Instrument/Spacecraft | Measurement Technique | Species | Energy range (MeV nucleon$^{-1}$) |
| ACE/ EPAM[a] | Residual Energy, E | H | ~0.04 – 5 |
| ACE/ULEIS | Time-of-Flight vs Residual Energy, E | H – Fe | ~0.1 – 14 |
| ACE/SIS | dE/dx vs E | He – Fe | ~5 – 170 |
| SoHO/ERNE | dE/dx vs E | H | ~2 – 140 |
| GOES/EPS[a] | dE/dx vs E | H | ~2 – 500 |
| SAMPEX/PET[a] | dE/dx vs E | H | ~19 – 500 |
| Notes: [a]Proton data from ACE/EPAM, GOES/EPS, and SAMPEX/PET during 6 large SEP events that were also associated with ground level enhancements (GLEs) are obtained from Mewaldt et al. (2012). | | | |

Table 2: Sampling intervals, spectral indices $\gamma_a$, $\gamma_b$ and break energies $E_B$ from the Band function fits for H, O & Fe, and power-law exponent, $\alpha$ of the Break energies vs. Q/M for 46 SEP events in this survey.

| Event No. (1) | Year (2) | Sampling Interval DOY, HHMM in UT (3) | Protons[a] | | | Oxygen[a,e] | | | Iron[a] | | | $\alpha$[a](13) |
|---|---|---|---|---|---|---|---|---|---|---|---|---|
| | | | $\gamma_a$(4) | $\gamma_b$(5) | $E_B$ (6) | $\gamma_a$ (7) | $\gamma_b$ (8) | $E_B$(9) | $\gamma_a$(10) | $\gamma_b$(11) | $E_B$(12) | |
| 1[b] | 1998 | 110, 1253 − 116, 0054 | 1.53±0.01 | 1.53±0.19 | 15.07±3.73 | ... | 7.46±5.15 | 6.45±1.29 | 0.99±0.40 | 0.99±0.01 | 5.00±1.84 | 0.81±0.33 |
| 2[b] | 1998 | 126, 0908 − 129, 0000 | 1.87±0.02 | 2.15±0.05 | 6.44±4.26 | 1.94±0.04 | 3.11±0.16 | 14.97±2.51 | 2.19±0.08 | 11.85±1.35 | 25.64±8.79 | ... |
| 3 | 1998 | 129, 0548 − 133, 0000 | 0.60±0.41 | 2.39±0.43 | 4.47±2.48 | 0.91±0.05 | 2.60±0.05 | 4.01±0.37 | 1.24±0.11 | 2.81±0.21 | 7.31±1.88 | ... |
| 4 | 1998 | 310, 0012 − 316, 0000 | 0.98±0.75 | 4.42±1.86 | 1.80±0.90 | 0.24±0.15 | 4.11±0.10 | 0.31±0.03 | 1.27±0.11 | 3.92±0.11 | 0.44±0.06 | ... |
| 5 | 1999 | 21, 0117 − 22, 1439 | ... | ... | 0.87±0.17 | 0.14±0.09 | 3.28±0.04 | 0.35±0.03 | 0.72±0.31 | 2.60±0.04 | 0.27±0.08 | 0.53±0.31 |
| 6 | 1999 | 114, 1718 − 116, 1550 | 0.64±0.07 | ... | 7.34±1.51 | 0.92±0.05 | 3.16±0.06 | 1.40±0.11 | 0.23±0.11 | 3.12±0.07 | 0.59±0.07 | 1.43±0.21 |
| 7 | 1999 | 152, 2018 − 155, 0839 | 0.91±0.09 | 4.25±4.13 | 10.69±1.00 | 1.45±0.03 | 2.60±0.11 | 11.04±1.50 | 1.81±0.02 | 1.78±0.11 | 81.71±16.57 | ... |
| 8 | 1999 | 155, 0838 − 159, 1920 | 1.18±0.10 | ... | 9.32±2.52 | 0.97±0.05 | 3.58±0.07 | 1.18±0.09 | 0.47±0.08 | 3.72±0.08 | 0.45±0.03 | 1.79±0.20 |
| 9 | 2000 | 204, 1408 − 205, 2015 | 1.04±0.02 | 3.43±1.22 | 5.23±0.60 | 0.99±0.33 | 2.82±0.07 | 0.42±0.16 | 1.62±0.08 | 3.57±0.27 | 1.95±0.55 | ... |
| 10 | 2000 | 256, 1432 − 260, 1359 | 1.20±0.03 | ... | 5.92±1.20 | 1.06±0.02 | 4.96±0.14 | 2.16±0.08 | 1.01±0.03 | 3.75±0.07 | 0.91±0.06 | 1.22±0.23 |
| 11 | 2000 | 290, 0923 − 294, 1739 | 0.56±0.07 | 2.66±0.11 | 4.64±0.39 | 0.77±0.06 | 2.79±0.10 | 2.85±0.38 | 0.74±0.08 | 3.25±0.21 | 3.29±0.54 | 0.28±0.19 |
| 12 | 2000 | 299, 1347 − 302, 0229 | 2.14±0.04 | ... | 11.42±1.73 | 3.12±0.11 | 4.15±1.03 | 26.05±12.80 | 2.81±0.15 | 5.49±1.50 | 8.43±1.93 | 0.81±0.75[d] |
| 13 | 2001 | 28, 2213 − 32, 0445 | 1.00±0.02 | 4.15±0.79 | 2.29±0.10 | 0.47±0.07 | 3.79±0.06 | 0.47±0.03 | 0.47±0.06 | 3.26±0.03 | 0.31±0.02 | 1.26±0.18 |
| 14[b] | 2001 | 105,1432 − 108, 0400 | 1.19±0.02 | 2.15±0.07 | 11.73±1.23 | 0.78±0.07 | 2.66±0.05 | 1.42±0.20 | 0.59±0.12 | 2.62±0.04 | 0.62±0.10 | 1.95±0.26 |
| 15[b] | 2001 | 108, 0318 − 111, 2150 | 1.38±0.02 | 2.80±0.32 | 31.37±3.43 | 1.51±0.04 | 3.54±0.22 | 6.41±0.85 | 1.40±0.07 | 2.77±0.13 | 3.20±0.93 | 1.52±0.22 |
| 16[b] | 2001 | 360, 0548 − 362, 1800 | 1.45±0.03 | 2.95±0.06 | 20.75±2.31 | 0.66±0.09 | 2.72±0.08 | 1.22±0.22 | 0.26±0.20 | 2.38±0.02 | 0.27±0.05 | 2.89±0.26 |
| 17 | 2001 | 364, 2245 − 7, 2329 | 1.85±0.04 | ... | 17.00±1.32 | 1.54±0.05 | 5.95±1.67 | 5.58±0.61 | 1.33±0.03 | 5.79±0.50 | 2.74±0.14 | 1.12±0.11 |
| 18 | 2002 | 10, 1920 − 14, 0020 | 1.47±0.03 | 12.06±0.65 | 9.58±1.37 | 1.72±0.02 | 6.71±0.82 | 3.49±0.12 | 1.80±0.09 | 3.07±0.26 | 2.38±1.00 | 1.12±0.16 |
| 19 | 2002 | 51, 0648 − 55, 1200 | 1.49±0.11 | 2.94±0.30 | 6.86±1.66 | 2.15±0.04 | 3.20±0.18 | 12.72±2.18 | 2.13±0.04 | 3.29±0.19 | 7.01±1.59 | ... |
| 20 | 2002 | 188, 1243 − 191, 1214 | 1.60±0.10 | 3.22±1.43 | ... | 2.75±0.10 | ... | ... | 3.06±0.15 | ... | ... | ... |
| 21[b] | 2002 | 226, 0213 − 228, 1200 | 1.21±0.05 | ... | 4.71±0.29 | 1.24±0.06 | 3.95±0.11 | 1.15±0.14 | 1.61±0.07 | 3.84±0.12 | 0.95±0.14 | 1.29±0.17 |
| 22[b] | 2002 | 235, 2351 − 240, 2200 | 1.10±0.02 | 2.87±0.03 | 9.98±0.37 | 1.19±0.03 | 3.77±0.14 | 3.27±0.27 | 0.71±0.05 | 3.00±0.05 | 0.89±0.07 | 1.58±0.18 |
| 23 | 2003 | 151, 0523 − 153, 0205 | 1.27±0.01 | 3.43±0.44 | 3.87±0.19 | 1.68±0.03 | 3.60±0.06 | 1.37±0.13 | 1.44±0.06 | 3.42±0.07 | 0.99±0.14 | 0.98±0.13 |
| 24 | 2003 | 169, 0308 − 174, 2319 | 1.34±0.03 | ... | 4.33±0.48 | 1.20±0.04 | 5.07±0.34 | 1.85±0.16 | 1.06±0.05 | 5.10±0.41 | 0.94±0.09 | 0.99±0.17 |
| 25[b] | 2003 | 308, 2102 − 313, 1524 | 1.41±0.03 | 4.99±0.31 | 20.06±1.04 | 1.01±0.08 | 3.73±0.17 | 1.80±0.31 | 0.36±0.14 | 3.72±0.09 | 0.35±0.04 | 2.46±0.24 |
| 26 | 2003 | 336, 1208 − 340, 0300 | 0.26±0.09 | 5.26±0.88 | 2.52±0.16 | 0.51±0.04 | 5.01±0.14 | 1.12±0.06 | 0.46±0.07 | 4.69±0.21 | 0.44±0.03 | 1.15±0.13 |
| 27 | 2004 | 257, 1200 − 262, 1200 | 1.00±0.02 | 6.01±3.71 | 3.63±0.12 | 1.20±0.04 | 5.19±0.23 | 1.99±0.13 | 1.12±0.06 | 5.01±0.26 | 0.88±0.10 | 0.76±0.11 |
| 28 | 2004 | 263, 2128 − 268, 1314 | 1.00±0.02 | 2.83±0.29 | 2.47±0.21 | 1.74±0.11 | 3.40±0.04 | 0.49±0.08 | 2.24±0.10 | 3.68±0.17 | 1.07±0.27 | 1.15±0.39 |
| 29 | 2004 | 306, 0430 − 309, 1200 | 0.81±0.04 | ... | 14.42±1.30 | 2.18±0.05 | ... | ... | 1.81±0.08 | 2.89±0.15 | 4.39±1.73 | ... |
| 30 | 2005 | 167, 2048 − 170, 0000 | 1.39±0.15 | 0.48±0.45 | 157.22±30.26 | 2.09±0.02 | ... | ... | 1.80±0.05 | ... | 63.56±25.38 | ... |
| 31 | 2010 | 226, 1148 − 230, 0000 | 1.00±0.02 | 2.85±0.31 | 3.01±0.24 | 2.41±0.05 | 1.52±0.18 | 13.17±4.87 | 1.71±0.15 | 28.89±5.17 | 1.04±0.34 | ... |
| 32 | 2010 | 230, 0848 − 236, 0000 | 1.00±0.02 | 3.24±0.21 | 2.65±0.10 | 0.91±0.09 | 3.39±0.05 | 0.26±0.02 | 1.98±0.24 | 3.29±0.21 | 0.41±0.16 | 2.16±0.29[e] |
| 33 | 2011 | 66, 2112 − 72, 0000 | 1.56±0.04 | ... | 10.17±0.80 | 1.21±0.08 | 3.89±0.27 | 2.13±0.38 | 1.12±0.08 | 3.97±0.19 | 0.81±0.10 | 1.48±0.19 |
| 34 | 2011 | 158, 0747 – 162, 1800 | 2.25±0.09 | … | … | 2.83±0.07 | … | … | 2.86±0.05 | 0.75±0.01 | … | … |
| 35 | 2011 | 221, 0848 – 224, 0000 | 1.00±0.03 | 2.68±0.14 | 2.51±0.14 | 0.70±0.08 | 3.37±0.07 | 0.58±0.05 | … | 2.91±0.05 | 0.26±0.03 | 1.46±0.10 |
| 36 | 2011 | 330, 0948 – 335, 1200 | 0.63±0.11 | 4.67±0.98 | 3.08±0.31 | 0.95±0.06 | 3.65±0.10 | 1.07±0.11 | 1.00±0.08 | 3.93±0.15 | 0.62±0.06 | 0.87±0.22 |
| 37 | 2012 | 73, 1611 – 77, 1800 | 1.30±0.03 | 4.43±0.17 | 22.19±2.58 | 1.54±0.03 | 3.58±0.15 | 4.26±0.42 | 1.39±0.01 | 3.50±0.08 | 3.53±0.20 | 1.22±0.34 |
| 38 | 2012 | 138, 0318 – 143, 1000 | 1.00±0.05 | 3.04±0.15 | 2.61±0.09 | 1.01±0.04 | 3.18±0.03 | 0.83±0.06 | 0.80±0.08 | 3.01±0.04 | 0.45±0.04 | 1.16±0.12 |
| 39 | 2012 | 147, 2247 – 151, 1200 | 0.70±0.02 | 6.16±5.86 | 4.24±0.13 | 0.71±0.07 | 4.56±0.33 | 1.24±0.13 | … | 3.88±0.27 | 0.37±0.04 | 1.59±0.11 |
| 40 | 2012 | 205, 0618 – 210, 0000 | 1.86±0.11 | 1.35±0.94 | 120.52±50.32 | 1.84±0.07 | … | 58.71±30.09 | 1.87±0.11 | 1.87±0.02 | 17.81±12.16 | 1.04±0.60 |
| 41 | 2013 | 101, 0848 – 105, 1200 | 1.83±0.04 | … | 129.77±92.05 | 2.73±0.07 | … | … | 2.54±0.04 | … | … | … |
| 42 | 2013 | 179, 0348 – 184, 0000 | 1.00±0.01 | 4.29±0.68 | 1.67±0.07 | 1.56±0.04 | 3.70±0.06 | 0.69±0.05 | 1.28±0.14 | 3.22±0.16 | 0.63±0.14 | 0.90±0.13 |
| 43 | 2013 | 229, 2048 – 232, 2259 | … | 3.22±0.49 | 0.69±0.14 | … | 3.57±0.05 | 0.14±0.01 | … | 3.35±0.16 | 0.10±0.05 | 0.71±0.52[e] |
| 44 | 2013 | 362, 1400 – 365, 0000 | 1.53±0.06 | 3.33±0.32 | 37.85±9.45 | 2.35±0.03 | … | 104.82±98.69 | 1.10±0.11 | 2.69±0.10 | 1.56±0.42 | 1.65±1.35[f] |
| 45 | 2014 | 4, 2100 – 7, 1200 | 1.75±0.11 | 1.75±0.18 | 8.91±1.04 | 2.56±0.02 | … | … | 2.30±0.03 | 1.19±0.50 | … | … |
| 46 | 2014 | 50, 0100 – 54, 0000 | 1.34±0.06 | 2.89±0.01 | 1.05±0.1 | 1.47±0.09 | 4.15±0.12 | 0.62±0.09 | 1.53±0.14 | 4.29±0.32 | 0.44±0.08 | 0.51±0.13 |

Notes:

[a] Fit parameters with relative uncertainties >100% indicate poor fits to the data and have been eliminated from this Table and the analyses.

[b] Also included proton spectra from Mewaldt et al. (2012) study of 16 SEPs associated with ground level enhancements (GLEs) from solar cycle 23.

[c] O fit parameters, included here for completeness, are also provided in Table 3 of Paper 1 (Desai et al. 2015).

[d] $\alpha$ is obtained using break energies obtained from fits to the He-Fe fluence spectra between ~1–50 MeV nucleon$^{-1}$; limited energy range for ULEIS.

[e] $\alpha$ is obtained using break energies obtained from fits to the He-Fe fluence spectra between ~0.1–10 MeV nucleon$^{-1}$; limited energy range for SIS.

[f] He, C, N, Ne, Mg, Si, S, and Ca fluence spectra exhibit no evidence of steepening in the combined ULEIS and SIS energy ranges.

| Event No. | $^3$He/$^4$He (x $10^{-2}$) | Fe/O ratio (@MeV nucleon$^{-1}$) | |
|---|---|---|---|
| (1) | (2) | 0.16-0.23 (3) | 15-21 (4) |
| 21 | 1.682±0.094 | 0.323±0.007 | 0.123±0.023 |
| 28 | 0.208±0.047 | 0.244±0.005 | 0.109±0.045 |
| 32 | <0.054 | 0.140±0.004 | 0.053±0.039 |
| 37 | 0.171±0.043 | 0.234±0.005 | 0.252±0.024 |
| 42 | <0.062 | 0.112±0.004 | 0.281±0.133 |

Table 3: The 0.5-2 MeV nucleon$^{-1}$ $^3$He/$^4$He and Fe/O ratios at ~0.193 and ~18 MeV nucleon$^{-1}$ for 5 events with large deviations in the spectral breaks for Fe and Ca (see §3.5; taken from Paper 1).

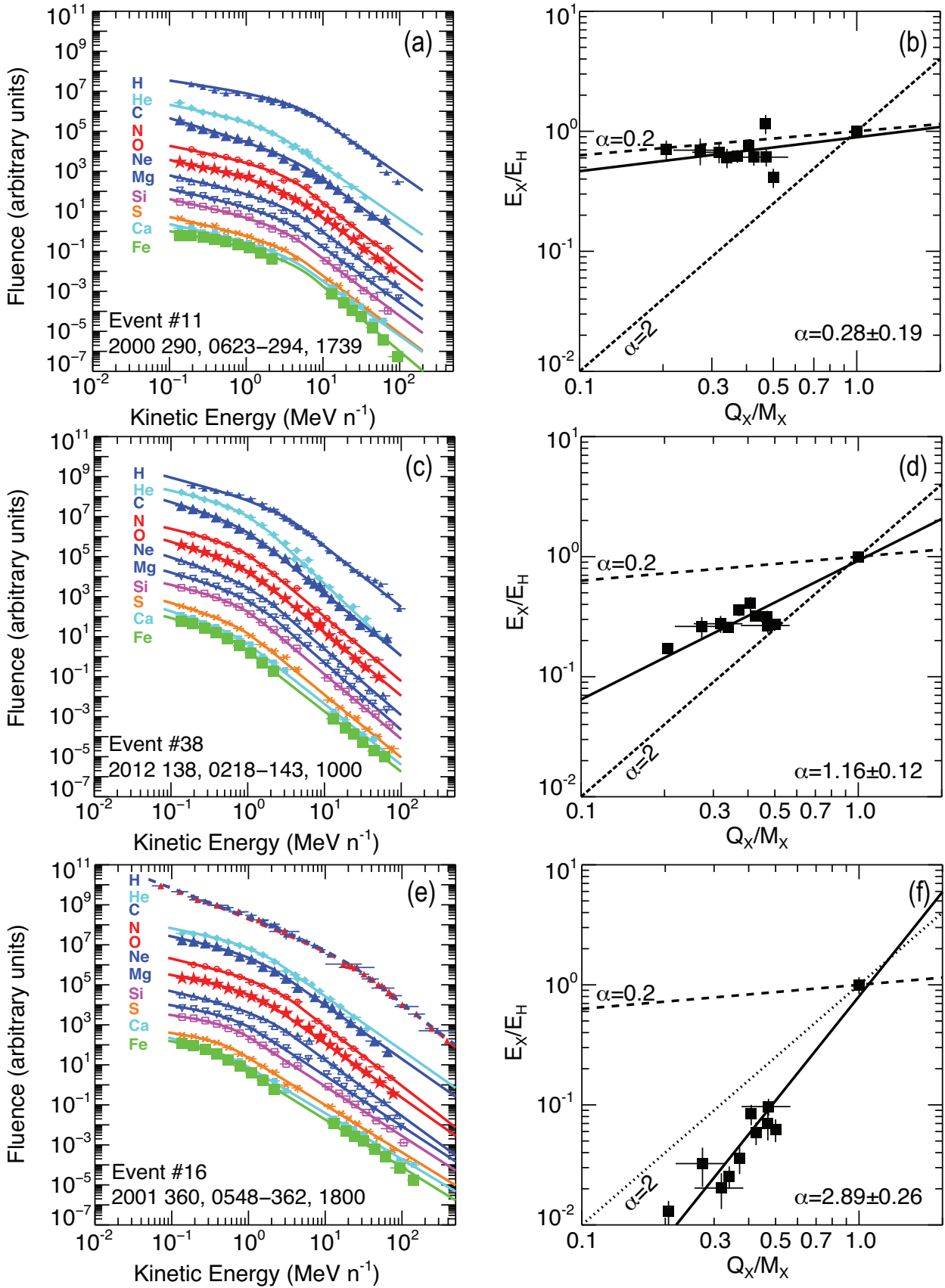

FIGURE 1:

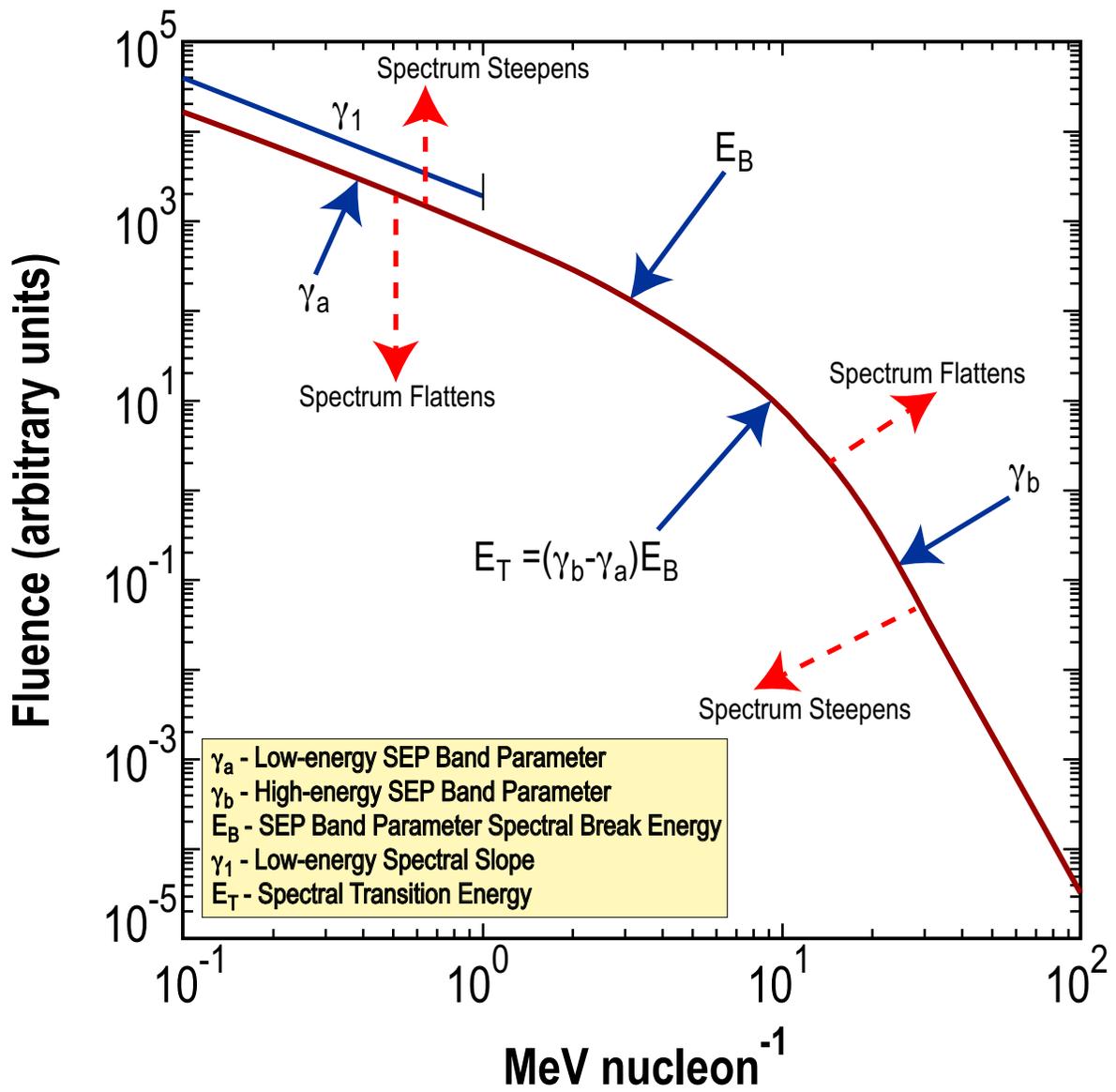

FIGURE 2:

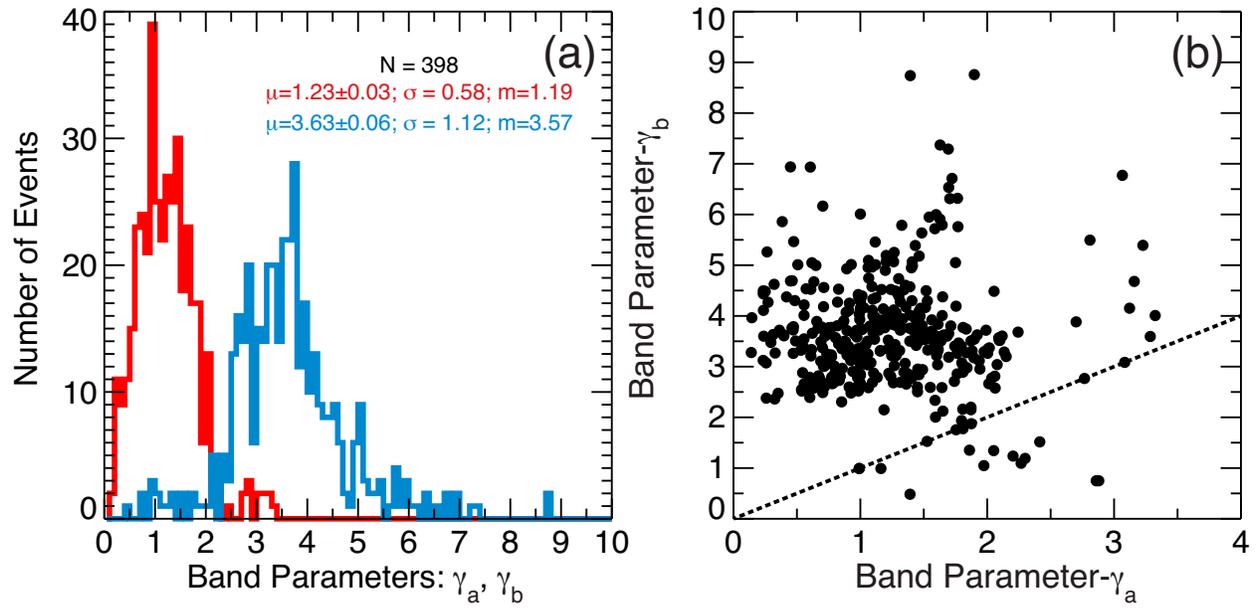

FIGURE 3:

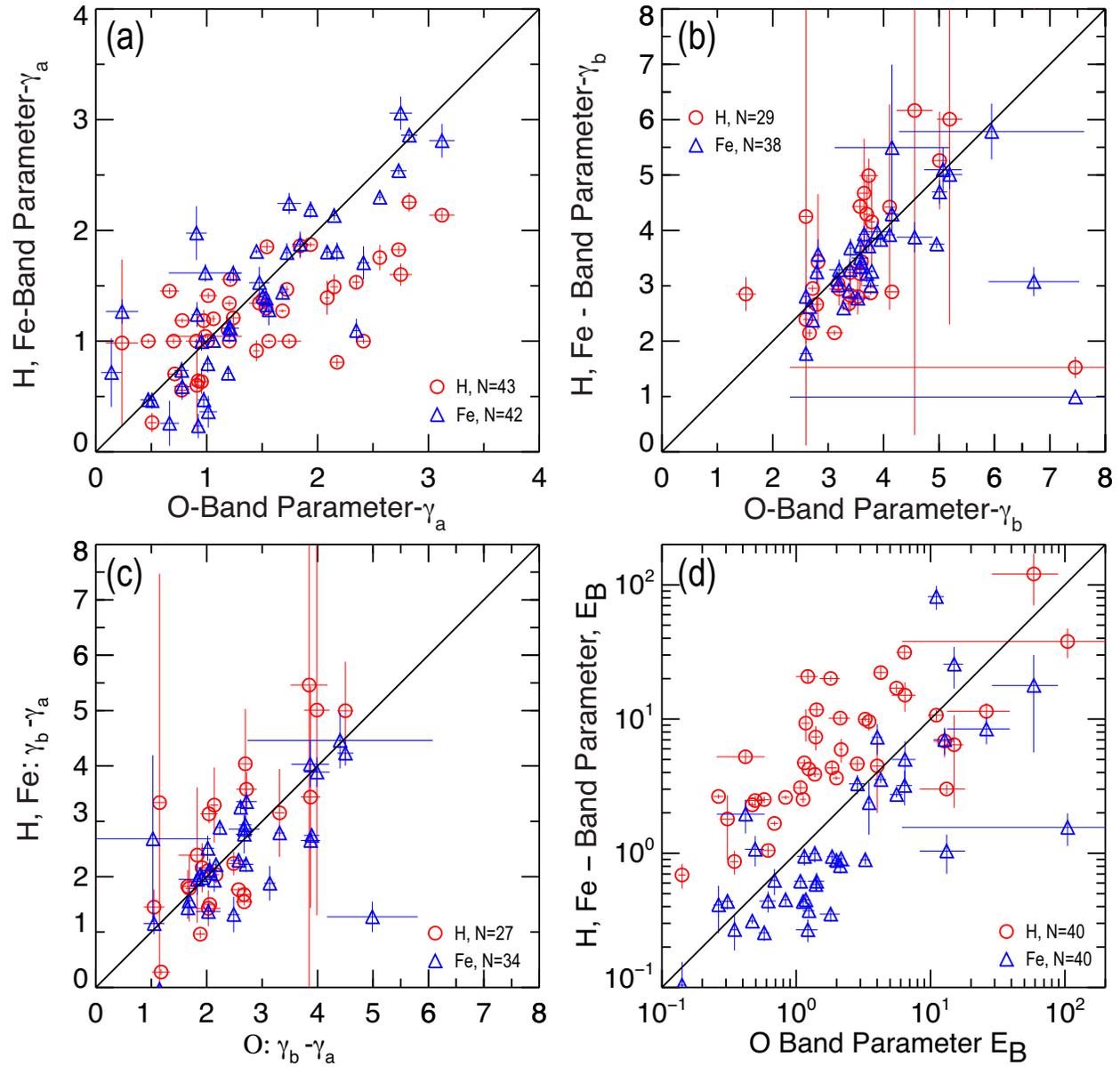

FIGURE 4:

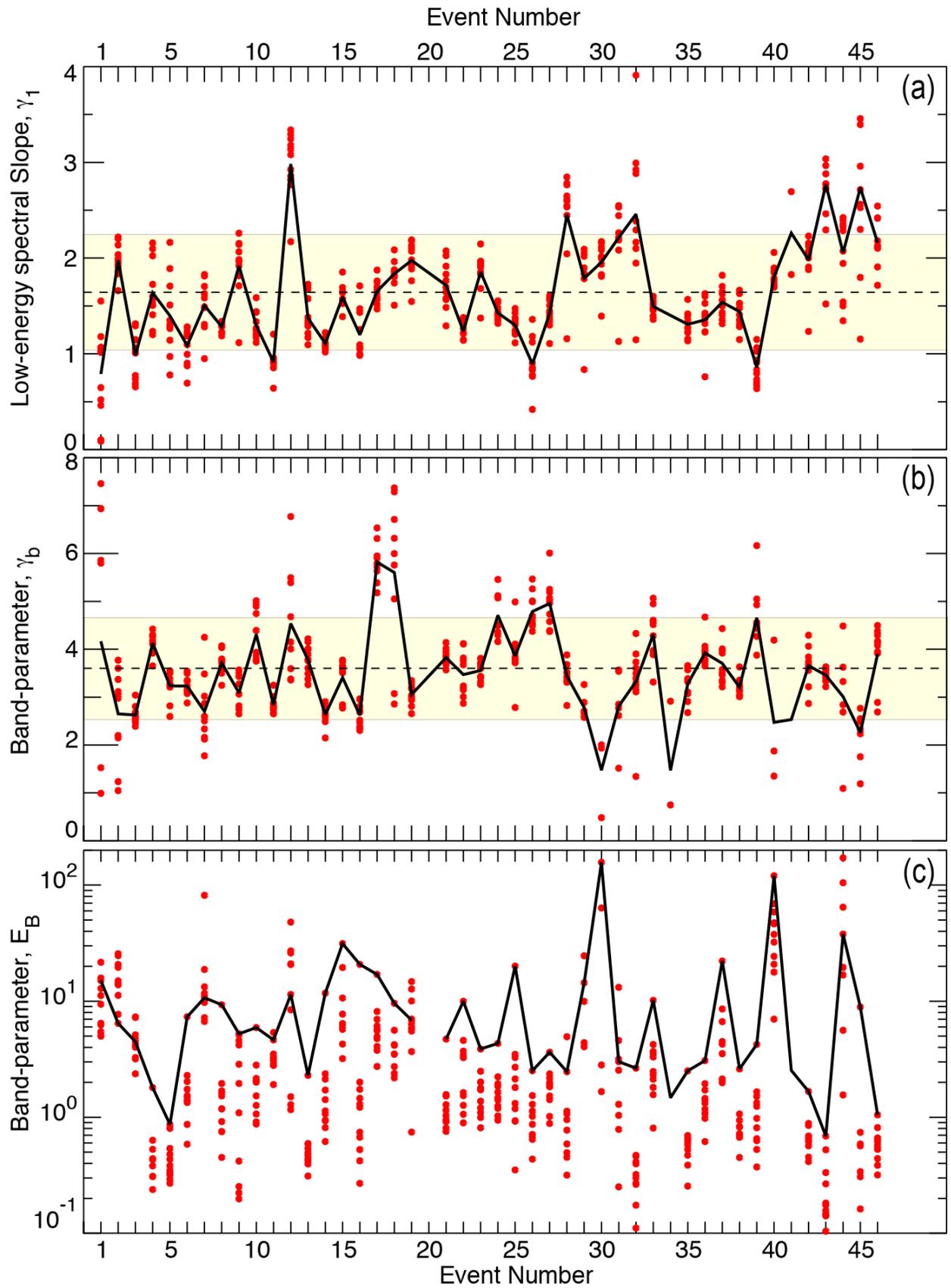

FIGURE 5:

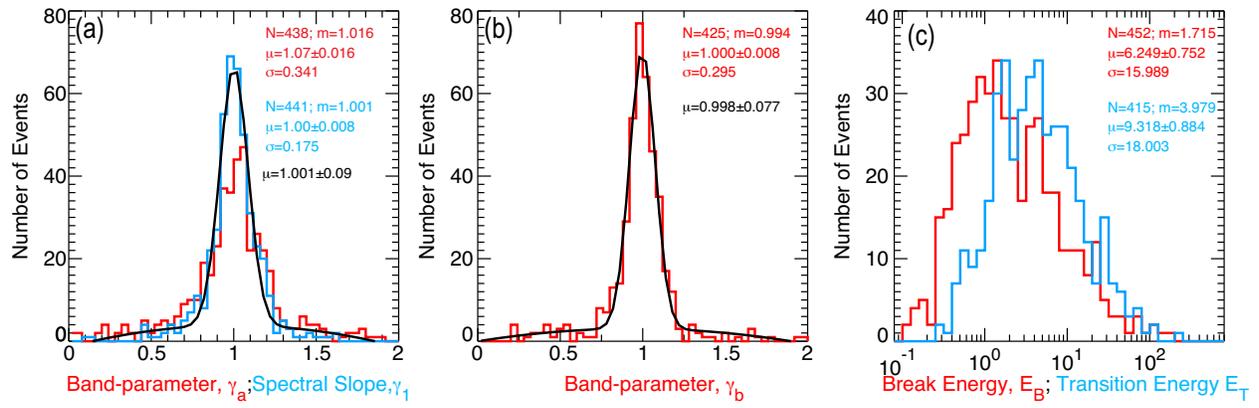

FIGURE 6:

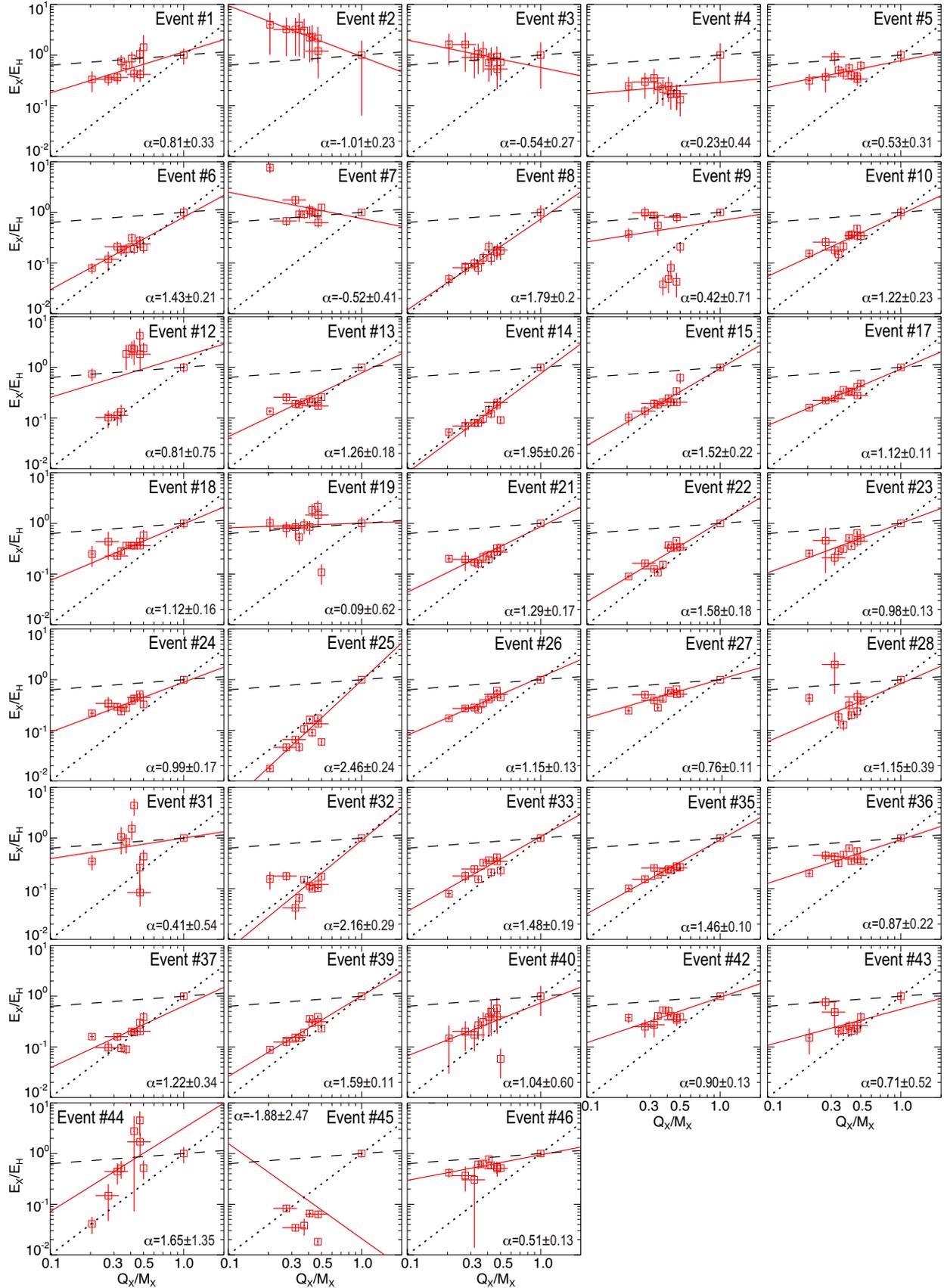

FIGURE 7:

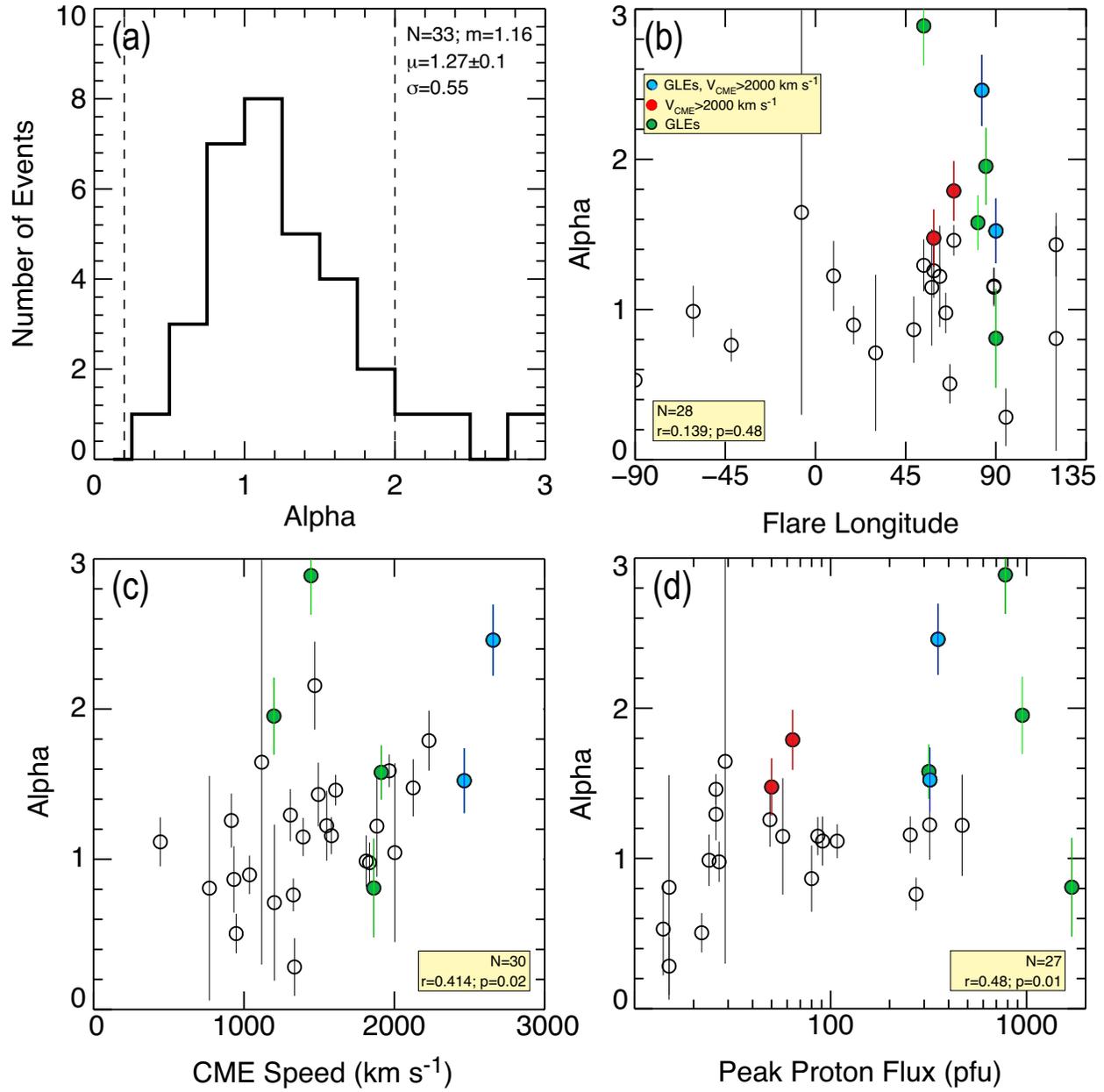

FIGURE 8:

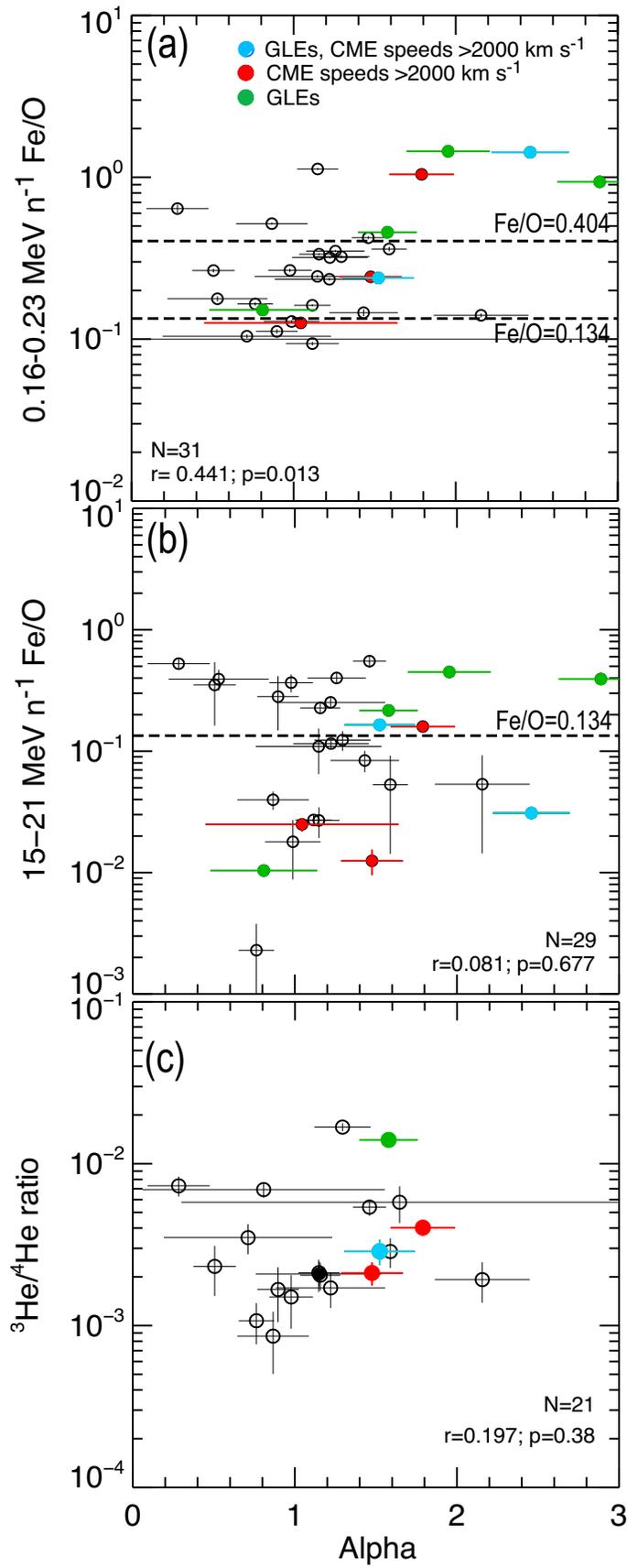

FIGURE 9:

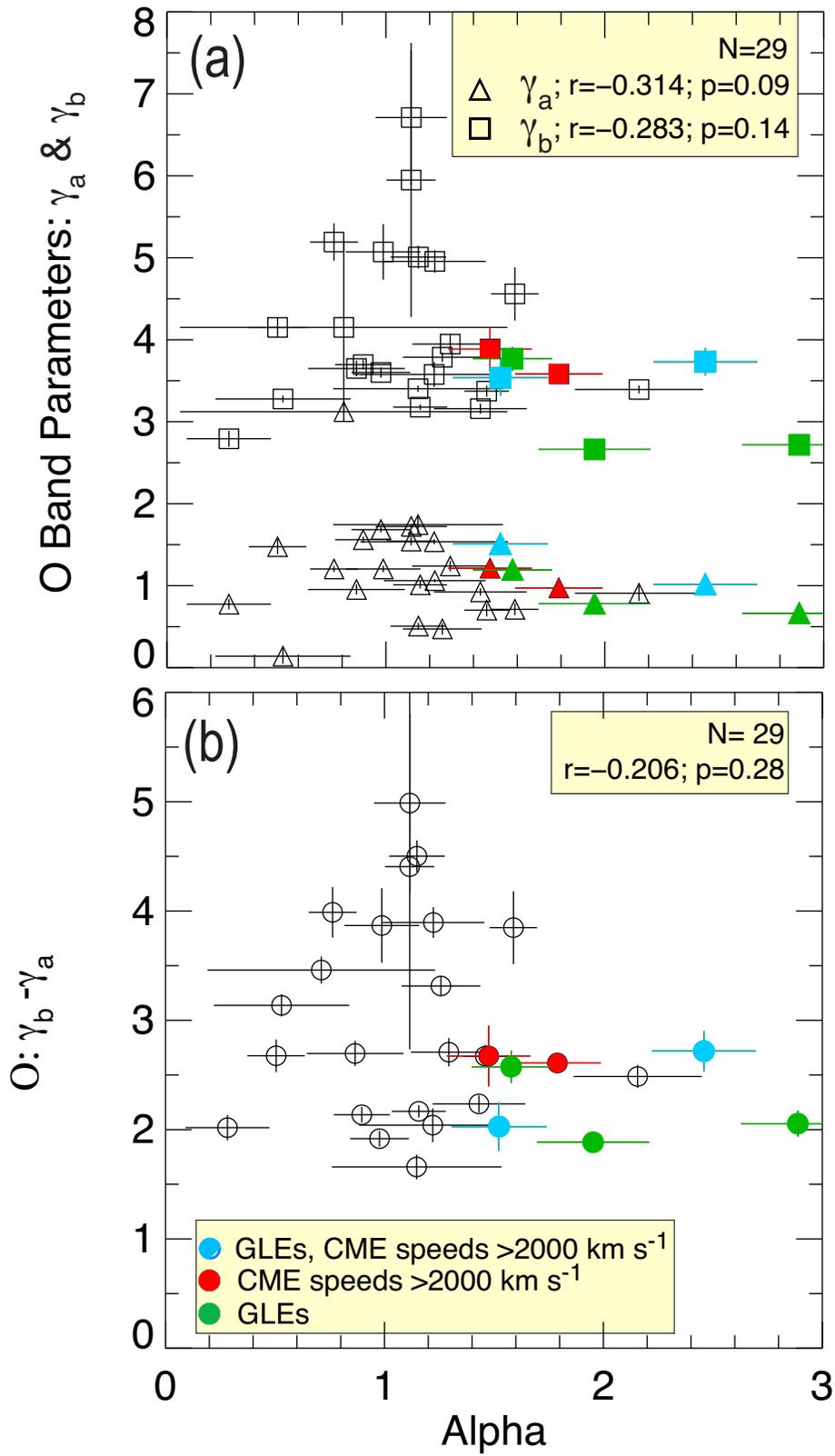

FIGURE 10:

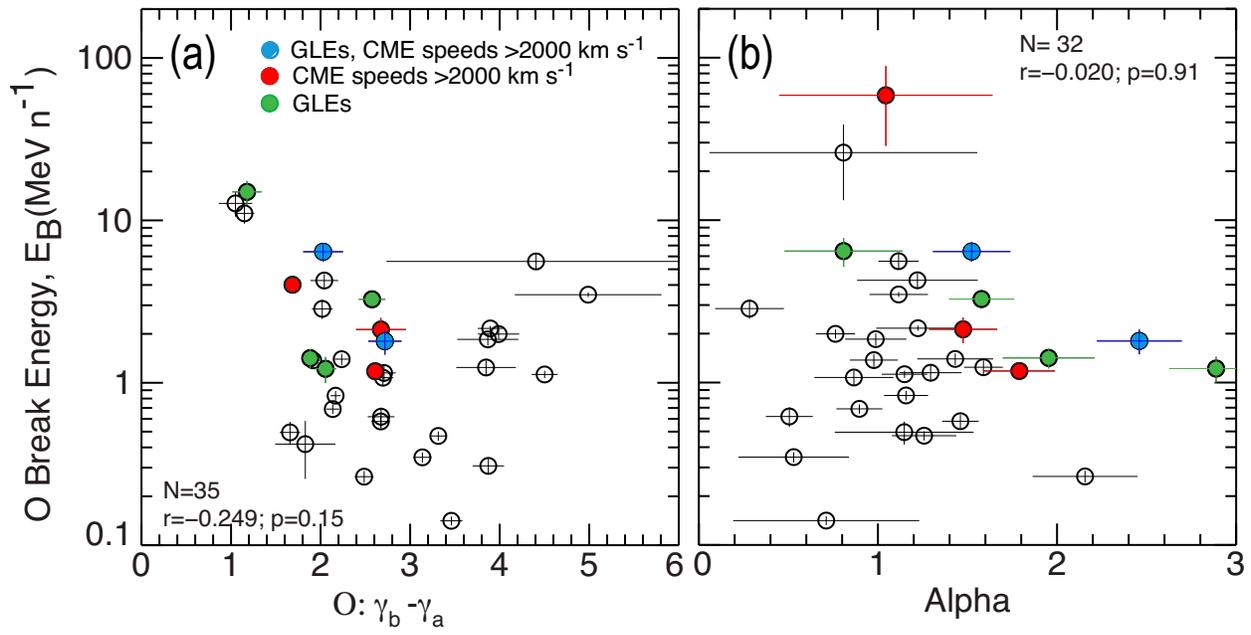

FIGURE 11:

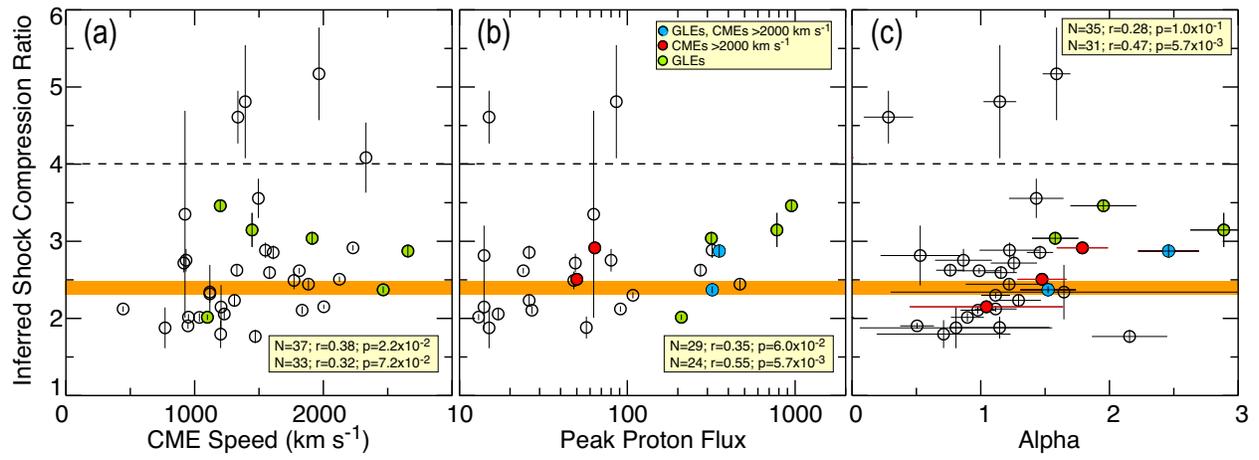

FIGURE 12:

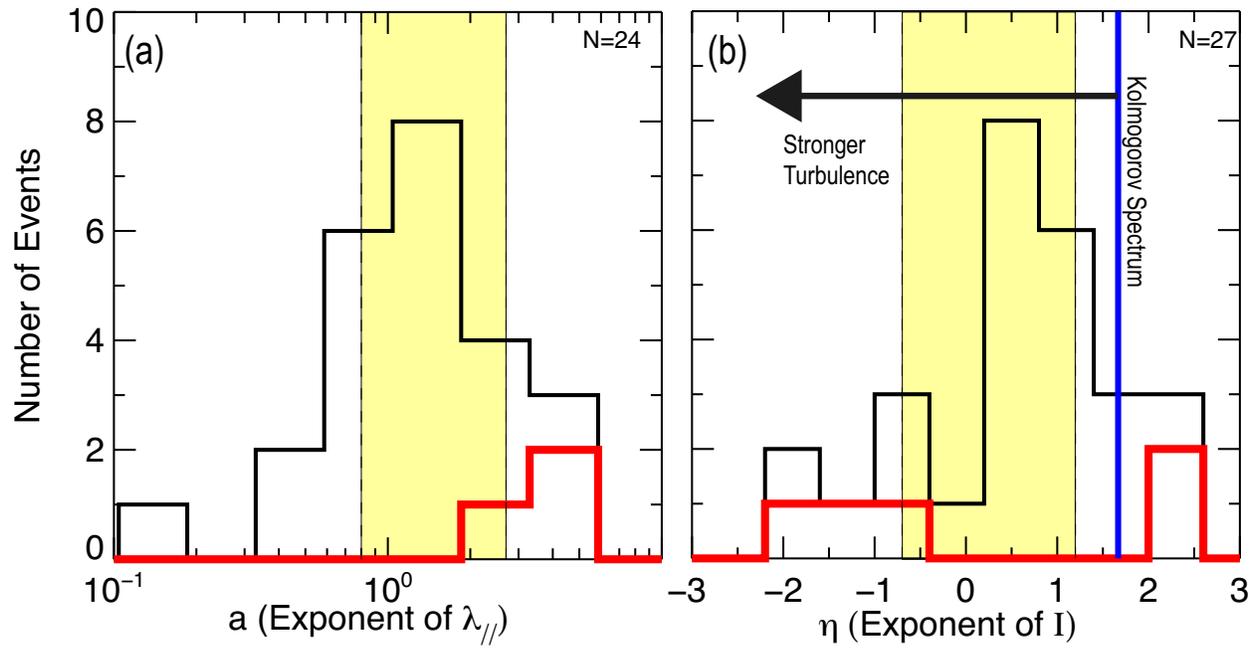

FIGURE 13: